\documentclass[11pt,a4paper,english,nofootinbib]{revtex4}
\usepackage{lmodern}

\usepackage[T1]{fontenc}
\usepackage[latin9]{inputenc}
\usepackage{color}
\usepackage{babel}
\usepackage{amsmath}
\usepackage{amssymb}
\usepackage{graphicx}
\usepackage{esint}
\usepackage[unicode=true,pdfusetitle,
 bookmarks=true,bookmarksnumbered=false,bookmarksopen=false,
 breaklinks=false,pdfborder={0 0 1},backref=false,colorlinks=false]
 {hyperref}

\makeatletter

\pdfpageheight\paperheight
\pdfpagewidth\paperwidth

\@ifundefined{textcolor}{}
{%
 \definecolor{BLACK}{gray}{0}
 \definecolor{WHITE}{gray}{1}
 \definecolor{RED}{rgb}{1,0,0}
 \definecolor{GREEN}{rgb}{0,1,0}
 \definecolor{BLUE}{rgb}{0,0,1}
 \definecolor{CYAN}{cmyk}{1,0,0,0}
 \definecolor{MAGENTA}{cmyk}{0,1,0,0}
 \definecolor{YELLOW}{cmyk}{0,0,1,0}
 }

\@ifundefined{definecolor}
 {\usepackage{color}}{}
\usepackage{latexsym}\usepackage{bm}

\makeatother

\begin{document}

\title{All unitary cubic curvature gravities in D dimensions}

\author{Tahsin Ça\u{g}r\i{} \c{S}i\c{s}man}

\email{sisman@metu.edu.tr}

\selectlanguage{english}%

\affiliation{Department of Physics,\\
 Middle East Technical University, 06531, Ankara, Turkey}

\author{\.{I}brahim Güllü }

\email{e075555@metu.edu.tr}

\selectlanguage{english}%

\affiliation{Department of Physics,\\
 Middle East Technical University, 06531, Ankara, Turkey}

\author{Bayram Tekin}

\email{btekin@metu.edu.tr}

\selectlanguage{english}%

\affiliation{Department of Physics,\\
 Middle East Technical University, 06531, Ankara, Turkey}

\date{\today}
\begin{abstract}
We construct all the unitary cubic curvature gravity theories built
on the contractions of the Riemann tensor in $D$-dimensional (anti)-de
Sitter spacetimes. Our construction is based on finding the equivalent
quadratic action for the general cubic curvature theory and imposing
ghost and tachyon freedom, which greatly simplifies the highly complicated
problem of finding the propagator of cubic curvature theories in constant
curvature backgrounds. To carry out the procedure we have also classified
all the unitary quadratic models. We use our general results to study
the recently found cubic curvature theories using different techniques
and the string generated cubic curvature gravity model. We also study
the scattering in critical gravity and give its cubic curvature extensions.\tableofcontents{}
\[
\]

\end{abstract}
\maketitle

\section{Introduction}

Recently, a new purely cubic curvature theory in $\left(4+1\right)$-dimensions
with physically attractive properties was found in two separate works
\cite{Oliva1,MyersRobinson}. In constructing this theory, different
guiding principles were used. For example, in \cite{Oliva1} the new
cubic theory was uniquely singled out from the requirement that, just
like in the higher curvature part of the $\left(2+1\right)$-dimensional
new massive gravity (NMG) \cite{BHT}, the trace of the field equations
be second order and proportional to the Lagrangian itself, and that
the field equations be second order for spherically symmetric solutions.
On the other hand, the authors of \cite{MyersRobinson} were searching
for a cubic curvature theory with analytical, simple, spherically
symmetric solutions and found the same theory, and coined it the {}``quasi-topological
gravity''. The theory becomes more interesting once lower powers
of curvature are added to it. $D$-dimensional extensions of the theory
and its black hole solutions were found in \cite{Oliva1,MyersRobinson,Oliva2,Mann}.
Also, extensive work on the model with regard to holography and the
AdS/CFT appeared in \cite{MyersPaulosSinha,MyersSinha1,Sinha2,MyersSinha2}.
Especially in \cite{MyersSinha1,MyersSinha2}, the trace anomaly structure
of the dual conformal field theory is studied and a (holographic)
$c$-theorem \emph{à la }\cite{Zamolodchikov} and \cite{Cardy} was
proposed for both even and odd dimensions.

In this work, we will construct \emph{all} the viable cubic curvature
gravity theories with the guiding principle that the theory be unitarity
around its (anti)-de Sitter {[}(A)dS{]} as well as its flat vacuum.
We will find the most general cubic curvature theories that are perturbatively
unitary in $D$ dimensions. To check the unitarity of a given gravity
theory around a fixed background, say, with metric $\bar{g}_{ab}$,
one has to find its propagator, that is the $O\left(h^{2}\right)$
expansion of the theory where $h_{ab}$ denotes the perturbation around
the background. This is in general a highly complicated problem even
for constant curvature backgrounds. But, the following observation
somewhat simplifies the analysis: In generic $D$ dimensions, if a
higher curvature gravity theory is to be unitary, its propagator should,
necessarily, reduce to the propagator of one of the following \textcolor{black}{four}
unitary gravity models (the details and the unitarity regions of these
theories will be given in Section II).\global\long\def\labelenumi{{\bf {\Roman{enumi}.}}}
 
\begin{enumerate}
\item The cosmological Einstein theory, $\frac{1}{\kappa}\left(R-2\Lambda_{0}\right)$,
with a massless spin-2 degree of freedom which is unitary for any
$\Lambda_{0}$ and for $\kappa>0$. (We take the mostly plus sign
convention.) 
\item Einstein plus Gauss-Bonnet (EGB) theory, $\frac{1}{\kappa}\left(R-2\Lambda_{0}\right)+\gamma\left(R_{abcd}^{2}-4R_{ab}^{2}+R^{2}\right)$,
with a massless spin-2 degree of freedom, but with an effective Newton's
constant $\frac{1}{\kappa_{e}}=\frac{1}{\kappa}+\frac{4\Lambda\left(D-3\right)\left(D-4\right)}{\left(D-1\right)\left(D-2\right)}\gamma$,
where $\Lambda$ is the effective cosmological constant whose reality
puts a constraint on the parameters. (Here and below, the effective
Newton's constant refers to the coefficient in front of the linearized
Einstein tensor.) For unitarity $\kappa_{e}>0$. 
\item Einstein plus scalar curvature theory, $\frac{1}{\kappa}\left(R-2\Lambda_{0}\right)+\alpha R^{2}$,
with a massless spin-2 degree of freedom and a massive spin-0 mode
with an effective Newton's constant $\frac{1}{\kappa_{e}}=\frac{1}{\kappa}+\frac{4\Lambda D}{D-2}\alpha$.
For unitarity, $\kappa_{e}>0$ and the mass of the scalar mode $m_{s}^{2}=\frac{D-2}{4\left(D-1\right)\alpha\kappa_{e}}-\frac{2\Lambda D}{\left(D-1\right)\left(D-2\right)}$
should satisfy $m_{s}^{2}>0$ in dS, and the Breitenlohner-Freedman
(BF) bound \cite{BF,Mezincescu} 
\[
m_{s}^{2}\ge\frac{D-1}{2\left(D-2\right)}\Lambda,
\]
 in AdS. 
\item Linear combination of the above three theories, $\frac{1}{\kappa}\left(R-2\Lambda_{0}\right)+\alpha R^{2}+\gamma\left(R_{abcd}^{2}-4R_{ab}^{2}+R^{2}\right)$,
with a massless spin-2 degree of freedom and a massive spin-0 mode
with an effective Newton's constant $\frac{1}{\kappa_{e}}=\frac{1}{\kappa}+\frac{4\Lambda D}{D-2}\alpha+\frac{4\Lambda\left(D-3\right)\left(D-4\right)}{\left(D-1\right)\left(D-2\right)}\gamma$.
For unitarity, $\kappa_{e}>0$ and the conditions on the mass of the
scalar mode given in item \textbf{III} with the new $\kappa_{e}$. 
\end{enumerate}
\global\long\def\labelenumi{\arabic{enumi}.}
 Note that we do not consider the purely quadratic theories, since
we would like to have the well-tested Einstein's gravity in the infrared
region. But, our analysis can simply be extended to such models by
considering $\kappa\rightarrow\infty$. Of course, the above models
are not unitary for generic values of the parameters (especially in
a curved background), there are constraints which we shall explore
in more detail by looking at the one-particle exchange amplitude between
two covariantly conserved sources \cite{GulluTekin}. We, also, do
not consider the Fierz-Pauli massive gravity (with mass $m^{2}\left(h_{\mu\nu}^{2}-h^{2}\right)$),
even though it is unitary, it is not gauge invariant, so its propagator
will not match any of the higher curvature models that we shall discuss.
The classification of all the unitary cubic curvature theories that
we shall present is exhaustive and devoid of any nonphysical or hard
to justify assumptions. We have done a similar classification in three
dimensions recently \cite{Gullu-Cubic3D} where we have found all
the bulk and boundary cubic curvature gravity theories.

\textcolor{black}{The approach of analyzing the unitarity of the cubic
curvature gravity theories by determining the quadratic curvature
action which has the same propagator with the original cubic curvature
action can also be used to construct the cubic curvature extensions
of the critical gravity theory found in \cite{LuPope,DeserLiu}. As
in the case of the unitary theories listed above, one can analyze
the propagator structure of the critical theory by taking the proper
limits of the one-particle exchange amplitude given in \cite{GulluTekin}
and reveal the double pole structure of the critical gravity implying
the logarithmic modes found in \cite{Alishahiha,GulluGurses,Rosseel}.}

The layout of the paper is as follows: In Section II, we discuss the
propagator and the spectra of quadratic curvature gravity theories
in (A)dS and find the unitary regions that lead us to the above classification.
In Section III, which is the bulk of this paper, we find the equivalent
quadratic action that has the same free theory of a generic cubic
curvature theory in $D$-dimensional (A)dS spacetime. In that section,
we also discuss the unitarity of the previously found cubic curvature
gravity theories mentioned above \cite{Oliva1,MyersRobinson,Oliva2,MyersSinha2}
and the cubic curvature theory generated as an effective theory in
the bosonic string theory \cite{Tseytlin}.\textcolor{black}{{} In Appendix
A, we give the one-particle exchange amplitude of the critical gravity
and construct its cubic curvature extensions. }Some details of the
computations regarding the equivalent quadratic actions are delegated
to the Appendix B.

\section{Propagator Structure of Quadratic Curvature Gravities\label{sec:Propagator}}

Any higher curvature theory built on the powers of the Riemann tensor
and its contractions in the form $f\left(R_{abcd}\right)$ (but not
its derivatives) has a propagator that has the \emph{same structure}
as the propagator of the following most general quadratic theory 
\begin{equation}
I=\int d^{D}x\,\sqrt{-g}\left[\frac{1}{\kappa}\left(R-2\Lambda_{0}\right)+\alpha R^{2}+\beta R_{ab}^{2}+\gamma\left(R_{abcd}^{2}-4R_{ab}^{2}+R^{2}\right)\right],\label{eq:General_quad_curv}
\end{equation}
 with possibly redefined parameters. Hence, to study the tree-level
unitarity of a generic higher curvature theory, it suffices to understand
the unitarity of (\ref{eq:General_quad_curv}): that is to find the
constraints on the five parameters. In flat space with $\Lambda_{0}=0=\Lambda$,
it is well-known that for nonzero $\beta$, there is a massive ghost
and the theory is not unitary \cite{Stelle}. (Note that Stelle's
original result was only for $D=4$, in fact, it is now known that
a nonzero $\beta$ is allowed in $D=3$ \cite{BHT}. For $D\ge4$,
Stelle's result is still valid in flat space.) For curved backgrounds,
the unitarity analysis is actually quite involved and was carried
out in \cite{GulluTekin}, where the tree-level scattering amplitude
between two background covariantly conserved sources was computed
in the linearized version of the theory (\ref{eq:General_quad_curv})
augmented with a Fierz-Pauli (FP) mass term for the graviton. The
FP mass helps to fix the gauge, and as long as the background is curved,
after finding the propagator, it can be set to zero without introducing
any van Dam-Veltman-Zakharov type discontinuity.

Since our subsequent discussions of the cubic curvature theories rest
on the unitarity of the above quadratic theory, let us recapitulate
the relevant analysis on the unitarity of (\ref{eq:General_quad_curv}).
First of all, the two maximally symmetric vacua, with the Riemann
tensor 
\begin{equation}
\bar{R}_{abcd}=\frac{2\Lambda}{\left(D-1\right)\left(D-2\right)}\left(\bar{g}_{ac}\bar{g}_{bd}-\bar{g}_{ad}\bar{g}_{bc}\right),\label{eq:Maximally_symmetric}
\end{equation}
 of this theory satisfy 
\begin{equation}
\frac{\Lambda-\Lambda_{0}}{2\kappa}+f\Lambda^{2}=0,\qquad\qquad f\equiv\left(D\alpha+\beta\right)\frac{\left(D-4\right)}{\left(D-2\right)^{2}}+\gamma\frac{\left(D-3\right)\left(D-4\right)}{\left(D-1\right)\left(D-2\right)}.
\end{equation}
 One should keep in mind that $\gamma$ enters into the picture only
for $D\ge5$. The linearized field equations for the metric fluctuation
$h_{ab}\equiv g_{ab}-\bar{g}_{ab}$ around one of these constant curvature
backgrounds become \cite{DeserTekin,DeserTekinPRL} 
\begin{align}
T_{ab}\left(h\right) & =\frac{1}{\kappa_{e}}\mathcal{G}_{ab}^{L}+\left(2\alpha+\beta\right)\left(\bar{g}_{ab}\bar{\square}-\bar{\nabla}_{a}\bar{\nabla}_{b}+\frac{2\Lambda}{D-2}\bar{g}_{ab}\right)R^{L}+\beta\left(\bar{\square}\mathcal{G}_{ab}^{L}-\frac{2\Lambda}{D-1}\bar{g}_{ab}R^{L}\right),\label{eq:Linearized_eom}
\end{align}
 where we have defined the effective Newton's constant as 
\begin{equation}
\frac{1}{\kappa_{e}}\equiv\frac{1}{\kappa}+\frac{4\Lambda D}{D-2}\alpha+\frac{4\Lambda}{D-1}\beta+\frac{4\Lambda\left(D-3\right)\left(D-4\right)}{\left(D-1\right)\left(D-2\right)}\gamma.
\end{equation}
 This expression was used in the list of the unitary theories given
in the introduction. Here, $T_{ab}\left(h\right)$ contains the matter
source $\tau_{ab}$ and all the higher order terms $O\left(h^{1+n}\right)$.
The linearized Einstein tensor, the Ricci tensor and the scalar curvature
read 
\begin{equation}
\mathcal{G}_{ab}^{L}=R_{ab}^{L}-\frac{1}{2}\bar{g}_{ab}R^{L}-\frac{2\Lambda}{D-2}h_{ab},
\end{equation}
 
\begin{equation}
R_{ab}^{L}=\frac{1}{2}\left(\bar{\nabla}^{c}\bar{\nabla}_{a}h_{bc}+\bar{\nabla}^{c}\bar{\nabla}_{b}h_{ac}-\bar{\square}h_{ab}-\bar{\nabla}_{a}\bar{\nabla}_{b}h\right),\qquad R^{L}=-\bar{\square}h+\bar{\nabla}^{a}\bar{\nabla}^{b}h_{ab}-\frac{2\Lambda}{D-2}h.\label{eq:Lin_tensors}
\end{equation}
 It is important to realize that for all higher curvature gravity
models, the building blocks in (\ref{eq:Linearized_eom}) will remain
the same. Only the coefficients will be affected. For example, the
$n^{\text{th}}$ order higher curvature terms such as $\eta R^{n}$,
$\eta\left(R_{ab}^{2}\right)^{n/2}$, \emph{etc.} will change $\frac{1}{\kappa_{e}}$
with an additional term proportional to $\eta\Lambda^{n-1}$, and
will shift the coefficients $\left(2\alpha+\beta\right)$ and $\beta$
with a term proportional to $\eta\Lambda^{n-2}$.

The tree-level scattering amplitude between two background covariantly
conserved, $\bar{\nabla}_{a}T^{ab}=0$, sources, as shown in Figure,
is 
\begin{figure}
\includegraphics[scale=0.8]{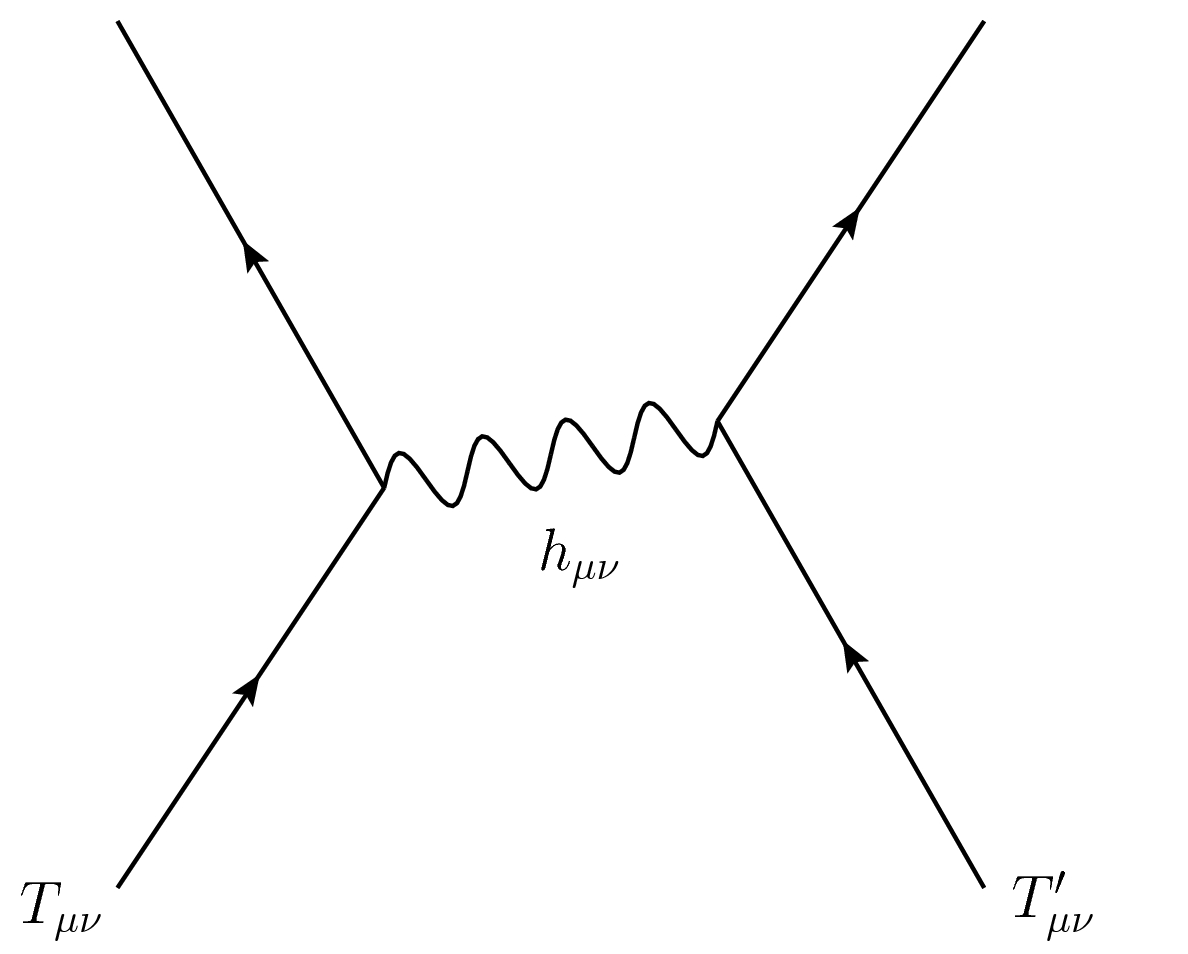}

\caption{Tree-level scattering amplitude between two background covariantly
conserved sources via the exchange of a graviton}
\end{figure}

\begin{equation}
A=\int d^{D}x\:\sqrt{-\bar{g}}T_{ab}^{\prime}\left(x\right)h^{ab}\left(x\right).\label{eq:Amplitude}
\end{equation}
 Here, $h_{ab}$ is the perturbation of the background metric created
by one of the sources which is represented with $T_{ab}$. Our normalization
of the amplitude is in such a way that in four dimensions for $\kappa=8\pi G_{N}$
we have the usual Einstein equation. The one-graviton scattering amplitude
between the two covariantly conserved sources $T_{ab}$ and $T_{ab}^{\prime}$
was calculated in \cite{GulluTekin} as 
\begin{align}
\frac{A}{\kappa_{e}}= & 2T_{ab}^{\prime}\left[\left(\kappa_{e}\beta\bar{\square}+1\right)\left(\triangle_{L}^{(2)}-\frac{4\Lambda}{D-2}\right)\right]^{-1}T^{ab}\nonumber \\
 & +\frac{2}{D-1}T^{\prime}\left[\left(\kappa_{e}\beta\bar{\square}+1\right)\left(\bar{\square}+\frac{4\Lambda}{D-2}\right)\right]^{-1}T\nonumber \\
 & -\frac{4\Lambda}{\left(D-1\right)^{2}\left(D-2\right)}T^{\prime}\left[\left(\kappa_{e}\beta\bar{\square}+1\right)\left(\bar{\square}+\frac{4\Lambda}{D-2}\right)\right]^{-1}\left[\bar{\square}+\frac{2\Lambda D}{\left(D-1\right)\left(D-2\right)}\right]^{-1}T\nonumber \\
 & -\frac{2}{\left(D-1\right)\left(D-2\right)}T^{\prime}\left[\kappa_{e}c\bar{\square}-1+\frac{2\Lambda D\kappa_{e}\left(c-\beta\right)}{\left(D-1\right)\left(D-2\right)}\right]^{-1}\left[\bar{\square}+\frac{2\Lambda D}{\left(D-1\right)\left(D-2\right)}\right]^{-1}T.\label{eq:One-particle_amp}
\end{align}
 where $c\equiv\frac{4\left(D-1\right)\alpha+D\beta}{D-2}$ and $\triangle_{L}^{(2)}$
is the Lichnerowicz operator acting on a symmetric rank-2 tensor.
To get (\ref{eq:One-particle_amp}), we have set the FP mass to zero
as discussed above. To simplify the notation we have omitted the integral
sign and the measure, and represented the Green's function as the
inverse of the operator. Computation of (\ref{eq:One-particle_amp})
is a somewhat cumbersome problem, since one has to find the tensor
Green's function corresponding to the equation $\mathcal{O}_{ab}^{\phantom{ab}cd}h_{cd}=T_{ab}$
where the operator $\mathcal{O}_{ab}^{\phantom{ab}cd}$ has fourth
order and second order derivatives, and a constant piece. The details
are laid out in \cite{GulluTekin}, therefore, we will not repeat
it here, but just note that the best way seems to be to decompose
$h_{ab}$ into the transverse and traceless parts as 
\begin{equation}
h_{ab}\equiv h_{ab}^{TT}+\bar{\nabla}_{(a}V_{b)}+\bar{\nabla}_{a}\bar{\nabla}_{b}\phi+\bar{g}_{ab}\psi,\label{eq:Decomposition}
\end{equation}
 then, gauge fix the equation with the help of a, say, FP mass term,
and decompose $T_{ab}$ in a similar way. This allows one to identify
the physical parts of $h_{ab}$, and relate them to their sources.
\textcolor{black}{As we noted above, unless $D=3$, $\beta$ introduces
a ghost.} This becomes more transparent if one organizes the $T-T^{\prime}$
part of (\ref{eq:One-particle_amp}) and rewrites it as 
\begin{align}
\frac{A}{\kappa_{e}}= & 2T_{ab}^{\prime}\left[\left(\kappa_{e}\beta\bar{\square}+1\right)\left(\triangle_{L}^{(2)}-\frac{4\Lambda}{D-2}\right)\right]^{-1}T^{ab}+\frac{2}{D-2}T^{\prime}\left[\left(\kappa_{e}\beta\bar{\square}+1\right)\left(\bar{\square}+\frac{4\Lambda}{D-2}\right)\right]^{-1}T\nonumber \\
 & -\frac{2\left(\beta+c\right)}{c\left(D-1\right)\left(D-2\right)}T^{\prime}\left[\left(\kappa_{e}\beta\bar{\square}+1\right)\left(\bar{\square}-m_{s}^{2}\right)\right]^{-1}T\\
 & +\frac{8\Lambda D\beta}{c\left(D-1\right)^{2}\left(D-2\right)^{2}}T^{\prime}\left[\left(\kappa_{e}\beta\bar{\square}+1\right)\left(\bar{\square}-m_{s}^{2}\right)\left(\bar{\square}+\frac{2\Lambda D}{\left(D-1\right)\left(D-2\right)}\right)\right]^{-1}T,\nonumber 
\end{align}
 where the mass of the scalar excitation is 
\begin{equation}
m_{s}^{2}=\frac{1}{c\kappa_{e}}-\frac{2\Lambda D}{\left(D-1\right)\left(D-2\right)}\left(1-\frac{\beta}{c}\right).\label{eq:Scalar_mass_with_beta}
\end{equation}
 If $\beta\ne0$ for generic $D$, the poles can be separated and
one can {}``see'' the ghost.\textcolor{black}{{} Since we have already
studied the $D=3$ theories in detail \cite{Gullu-Cubic3D}, we will
consider $D>3$.} Therefore, we first set $\beta=0$, then the apparent
pole $\bar{\square}=-\frac{2\Lambda D}{\left(D-1\right)\left(D-2\right)}$
decouples. It is still a complicated task to explore the unitary regions
of the remaining theory. One can somewhat simplify the analysis by
relying on the unitarity of the pure cosmological Einstein-Hilbert
theory whose unitarity has been checked with different techniques.
One can write the one-particle exchange amplitude of the Einstein's
gravity, that is $\alpha=\gamma=0$, as 
\begin{equation}
A=2\kappa\left[T_{ab}^{\prime}\left(\triangle_{L}^{(2)}-\frac{4\Lambda}{D-2}\right)^{-1}T^{ab}+\frac{1}{D-2}T^{\prime}\left(\bar{\square}+\frac{4\Lambda}{D-2}\right)^{-1}T\right],\label{eq:Amplitude_Einstein}
\end{equation}
 which represents a unitary interaction via a massless spin-2 graviton
for $\kappa>0$. This expression is our canonical example which will
guide us understand the spectrum and the unitarity of the other theories.
Note that this result reduces to that of \cite{Porrati} in four dimensions.

Among the unitary theories listed in the introduction, the EGB theory
augmented with the term $\alpha R^{2}$, which was classified as the
fourth theory, contains the first three cases in the proper limits,
therefore, let us start with the scattering amplitude of that theory
which reads 
\begin{align}
A=2\kappa_{e} & \Biggl[T_{ab}^{\prime}\left(\triangle_{L}^{(2)}-\frac{4\Lambda}{D-2}\right)^{-1}T^{ab}+\frac{1}{D-2}T^{\prime}\left(\bar{\square}+\frac{4\Lambda}{D-2}\right)^{-1}T,\nonumber \\
 & -\frac{1}{\left(D-1\right)\left(D-2\right)}T^{\prime}\left(\bar{\square}-m_{s}^{2}\right)^{-1}T\Biggr],\label{eq:Ampl_GB_and_R-aR2}
\end{align}
 where the mass of the spin-0 mode given in (\ref{eq:Scalar_mass_with_beta})
reduces to 
\begin{equation}
m_{s}^{2}=\frac{D-2}{4\left(D-1\right)\alpha\kappa_{e}}-\frac{2\Lambda D}{\left(D-1\right)\left(D-2\right)},\label{eq:scalar_mass}
\end{equation}
 and the effective Newton's constant is 
\begin{equation}
\frac{1}{\kappa_{e}}=\frac{1}{\kappa}+\frac{4\Lambda D}{D-2}\alpha+\frac{4\Lambda\left(D-3\right)\left(D-4\right)}{\left(D-1\right)\left(D-2\right)}\gamma.
\end{equation}
 Let us now study in some detail the unitarity regions and exploit
the constraints on the parameters of this theory for all values and
signs of $\Lambda$. For $\Lambda=0$, the amplitude reduces to the
simple flat space limit 
\begin{equation}
A=2\kappa\left[-T_{ab}^{\prime}\left(\partial^{2}\right)^{-1}T^{ab}+\frac{1}{D-2}T^{\prime}\left(\partial^{2}\right)^{-1}T-\frac{1}{\left(D-1\right)\left(D-2\right)}T^{\prime}\left(\partial^{2}-\frac{D-2}{4\left(D-1\right)\kappa\alpha}\right)^{-1}T\right],\label{eq:Flat_ampl_GB-R-aR2}
\end{equation}
 which is unitary for $\kappa>0$, and $m_{s}^{2}=\frac{D-2}{4\left(D-1\right)\kappa\alpha}>0$,
and hence, $\alpha>0$. As expected in flat space, the Gauss-Bonnet
term does not play any role at the linearized level, therefore, $\gamma$
is not restricted. On the other hand, in AdS ($\Lambda<0$), the mass
of the scalar excitation should satisfy the BF bound 
\begin{equation}
m_{s}^{2}\ge\frac{D-1}{2\left(D-2\right)}\Lambda,\label{eq:BF-bound}
\end{equation}
 which allows negative mass squared values. There is a wide region
of parameters that yield a unitary theory. To require \emph{attractive}
gravity or nonghost behavior, we impose $\kappa_{e}>0$, and we also
demand that in the limit of vanishing $\Lambda$ we have a unitary
theory, and hence require $\kappa>0$ and $\alpha>0$. {[}Note that
the latter two conditions can be removed to extend the unitarity regions.{]}
All these conditions restrict $\gamma$ to 
\begin{equation}
\gamma<-\left(\frac{1}{4\kappa\Lambda}+\frac{\alpha D}{D-2}\right)\frac{\left(D-1\right)\left(D-2\right)}{\left(D-3\right)\left(D-4\right)}.
\end{equation}
 In the dS ($\Lambda>0$) space, there is no BF bound on the scalar
mode, instead one has $m_{s}^{2}>0$, and the unitarity region reads

\begin{equation}
\gamma>-\left(\frac{1}{4\kappa\Lambda}+\frac{\alpha D\left(D-4\right)}{\left(D-2\right)^{2}}\right)\frac{\left(D-1\right)\left(D-2\right)}{\left(D-3\right)\left(D-4\right)}.
\end{equation}
 Note that $\Lambda$ is a function of $\Lambda_{0}$, $\alpha$ and
$\gamma$.

The scattering amplitude of the EGB theory, that is the second theory
in our list, is obtained by setting $\alpha=0$ in (\ref{eq:Ampl_GB_and_R-aR2})
and (\ref{eq:scalar_mass}) which freezes the scalar mode and reduces
the scattering amplitude to that of Einstein's gravity albeit with
a modified gravitational constant. To get the $R-2\Lambda_{0}+\alpha R^{2}$
theory, one can simply set $\gamma=0$ in (\ref{eq:Ampl_GB_and_R-aR2})
and (\ref{eq:scalar_mass}).

\textcolor{black}{One can also find the scattering amplitude for the
critical gravity by using (\ref{eq:One-particle_amp}) and the discussion
on this is given in Appendix A.}

\section{Equivalent Quadratic Action}

In the previous section, we have classified all the unitary quadratic
gravity theories in $D$-dimensional (A)dS spacetime. Our main task,
which is the purpose of this work, is now to extend this analysis
to cubic curvature theories. In $D$ dimensions, the most general
cubic curvature action constructed from the contractions of the Riemann
tensor, but not its derivatives, reads \cite{Fulling} 
\begin{align}
I=\int d^{D}x\sqrt{-g}\, & \Biggl[\frac{1}{\kappa}\left(R-2\Lambda_{0}\right)+\alpha R^{2}+\beta R_{ab}^{2}+\gamma\left(R_{abcd}^{2}-4R_{ab}^{2}+R^{2}\right)\nonumber \\
 & +\biggl(a_{1}R^{pqrs}R_{p\phantom{t}r}^{\phantom{p}t\phantom{r}u}R_{qust}+a_{2}R^{pqrs}R_{pq}^{\phantom{pq}tu}R_{rstu}+a_{3}R^{pq}R_{\phantom{rst}p}^{rst}R_{rstq}\label{eq:Cubic_action}\\
 & \phantom{+}+a_{4}RR^{pqrs}R_{pqrs}+a_{5}R^{pq}R^{rs}R_{prqs}+a_{6}R^{pq}R_{p}^{r}R_{qr}+a_{7}RR^{pq}R_{pq}+a_{8}R^{3}\Biggr)\Biggr].\nonumber 
\end{align}
 Altogether, we have 13 parameters and the main question is to find
the ranges of these parameters for which the theory is perturbatively
unitary around its (A)dS vacua (note that generically there will be
3 vacua). A direct way to answer this question is to compute the $O\left(h_{ab}^{2}\right)$
action and compare it with the unitary theories listed above. Computation
of the $O\left(h_{ab}^{2}\right)$ action directly by computing the
relevant tensors to $O\left(h_{ab}^{2}\right)$ is a highly cumbersome
analysis. To somewhat simplify the problem we will use a technique
developed in \cite{Hindawi} and extensively used in \cite{Gullu-BIUnitarity}
to study the unitarity of the Born-Infeld (BI) type gravities, and
in \cite{Gullu-Cubic3D} to show the unitarity of the BI extension
of NMG \cite{BINMG} and for the classification of all unitary cubic
curvature gravities in three dimensions including the cubic curvature
extension of NMG \cite{Sinha}. The technique consists of finding
an \emph{equivalent quadratic action} in curvature that has exactly
the same propagator as (\ref{eq:Cubic_action}). Moreover, as it will
be clear below, this equivalent quadratic action also has the same
vacua. This latter point is important since, with the help of the
equivalent quadratic action, one avoids the derivation of the full
nonlinear field equations to find the maximally symmetric vacua or
determine the effective cosmological constant in terms of the 13 parameters
of (\ref{eq:Cubic_action}). The procedure of how one finds the equivalent
quadratic action was described in \cite{Gullu-Cubic3D} which we briefly
repeat here. Consider a generic Lagrangian as a function of the Riemann
tensor, $\mathcal{L}\equiv\sqrt{-g}F\left(R_{cd}^{ab}\right)$. {[}We
will choose our independent field to be the up-up-down-down Riemann
tensor which will simplify the computations, since this choice does
not introduce the background metric or its inverse. In fact, in the
Appendix B, to further simplify the computations, we will consider
the scalar curvature $R$ and the Ricci tensor $R_{b}^{a}$ to be
the independent variables as well as the Riemann tensor $R_{cd}^{ab}$.{]}
Finding the vacua and the $O\left(h^{2}\right)$ action means computing
\begin{align}
\int d^{D}x\,\mathcal{L}\left(R_{cd}^{ab}\right)= & \int d^{D}x\,\mathcal{L}\left(\bar{R}_{cd}^{ab}\right)+\int d^{D}x\,\left[\frac{\delta\mathcal{L}}{\delta g^{ab}}\right]_{\bar{g}_{ab}}\delta g^{ab}\label{eq:Standard_expansion}\\
 & +\frac{1}{2}\int d^{D}x\,\delta g^{cd}\left[\frac{\delta\mathcal{L}}{\delta g^{cd}\delta g^{ab}}\right]_{\bar{g}_{ab}}\delta g^{ab}+\dots.\nonumber 
\end{align}
 The first term in the right-hand side is irrelevant for our purposes,
the second term gives the field equations and the third one gives
the propagator. Finding an equivalent quadratic action means one finds
a quadratic Lagrangian in curvature $\mathcal{L}_{\text{quad-equal}}\equiv\sqrt{-g}f_{\text{quad-equal}}\left(R_{cd}^{ab}\right)$
that has the same $O\left(h^{0}\right)$, $O\left(h\right)$ and $O\left(h^{2}\right)$
expansions as the original Lagrangian $\mathcal{L}$. The equivalent
quadratic Lagrangian can be obtained by expanding $F\left(R_{cd}^{ab}\right)$
around the \emph{yet to be found} constant curvature vacuum $\bar{R}_{cd}^{ab}$
given in (\ref{eq:Maximally_symmetric}) up to second order as 
\begin{equation}
f_{\text{quad-equal}}\left(R_{cd}^{ab}\right)\equiv\sum_{i=0}^{2}\left[\frac{\partial^{i}F}{\partial\left(R_{cd}^{ab}\right)^{i}}\right]_{\bar{R}_{cd}^{ab}}\left(R_{cd}^{ab}-\bar{R}_{cd}^{ab}\right)^{i}.\label{eq:f_quad}
\end{equation}
 Let us stress that $\sqrt{-g}f_{\text{quad-equal}}\left(R_{cd}^{ab}\right)$
has the same $O\left(h^{0}\right)$, $O\left(h\right)$ and $O\left(h^{2}\right)$
expansions as the original Lagrangian. This simple, yet remarkable,
result follows from the fact that all the, say, $O\left(h^{2}\right)$
terms of a given Lagrangian $\mathcal{L}\equiv\sqrt{-g}F\left(R_{cd}^{ab}\right)$
are in the form 
\begin{align}
\mathcal{L}_{O\left(h^{2}\right)}=\delta g^{cd}\left[\frac{\delta\mathcal{L}}{\delta g^{cd}\delta g^{ab}}\right]_{\bar{g}}\delta g^{ab}= & \left[\left(\sqrt{-g}\right)_{O\left(h^{0}\right)}+\left(\sqrt{-g}\right)_{O\left(h^{1}\right)}+\left(\sqrt{-g}\right)_{O\left(h^{2}\right)}\right]\nonumber \\
 & \times\left[F_{O\left(h^{0}\right)}+F_{O\left(h^{1}\right)}+F_{O\left(h^{2}\right)}\right]\nonumber \\
= & \left(\sqrt{-g}\right)_{O\left(h^{0}\right)}F_{O\left(h^{2}\right)}+\left(\sqrt{-g}\right)_{O\left(h^{1}\right)}F_{O\left(h^{1}\right)}+\left(\sqrt{-g}\right)_{O\left(h^{2}\right)}F_{O\left(h^{0}\right)},
\end{align}
 where $\left(\sqrt{-g}\right)_{O\left(h^{n}\right)}$ refers to the
$O\left(h^{n}\right)$ expansion of $\sqrt{-g}$, $F$ terms follow
similarly. Here, all the terms involved in $\left[F_{O\left(h^{0}\right)}+F_{O\left(h^{1}\right)}+F_{O\left(h^{2}\right)}\right]$
are given as 
\begin{align}
F_{O\left(h^{0}\right)}+F_{O\left(h^{1}\right)}+F_{O\left(h^{2}\right)}= & F\left(\bar{R}_{cd}^{ab}\right)+\left[\frac{\partial F}{\partial R_{cd}^{ab}}\right]_{\bar{R}_{cd}^{ab}}\left(R_{cd}^{ab}\right)_{O\left(h\right)}\nonumber \\
 & +\left\{ \left[\frac{\partial F}{\partial R_{cd}^{ab}}\right]_{\bar{R}_{cd}^{ab}}\left(R_{cd}^{ab}\right)_{O\left(h^{2}\right)}+\left[\frac{\partial^{2}F}{\partial\left(R_{cd}^{ab}\right)^{2}}\right]_{\bar{R}_{cd}^{ab}}\left[\left(R_{cd}^{ab}\right)_{O\left(h\right)}\right]^{2}\right\} .
\end{align}
 As it is clear from the second line of this expression, the linear
order in curvature also contributes to the $O\left(h^{2}\right)$
terms. Generically, all the terms up to order $n$ in the expansion
\begin{equation}
\sum_{i=1}^{n}\left[\frac{\partial^{i}F}{\partial\left(R_{cd}^{ab}\right)^{i}}\right]_{\bar{R}_{cd}^{ab}}\left(R_{cd}^{ab}-\bar{R}_{cd}^{ab}\right)^{i}
\end{equation}
 contribute to $F_{O\left(h^{n}\right)}$.%
\footnote{In fact, another way of finding the equivalent quadratic action is
to observe that at the desired order of $O\left(h^{2}\right)$, $\left(\text{Riem}-\bar{\text{Riem}}\right)^{3}=0$.
We thank a referee for pointing this fact.%
}

Let us now apply these general ideas to (\ref{eq:Cubic_action}).
First of all, we do not need to consider the constant, the linear
and the quadratic terms in the curvature, since they are already in
the desired form. Let us rewrite the cubic action in such a way that
$R_{cd}^{ab}$ appears as the only independent variable 
\begin{align}
F\left(R_{cd}^{ab}\right)\equiv & a_{1}R_{rs}^{pq}R_{pt}^{ru}R_{qu}^{st}+a_{2}R_{rs}^{pq}R_{pq}^{tu}R_{tu}^{rs}+a_{3}R_{q}^{p}R_{tp}^{rs}R_{rs}^{tq}+a_{4}RR_{rs}^{pq}R_{pq}^{rs}\nonumber \\
 & +a_{5}R_{q}^{p}R_{s}^{r}R_{pr}^{qs}+a_{6}R_{q}^{p}R_{p}^{r}R_{r}^{q}+a_{7}RR_{q}^{p}R_{p}^{q}+a_{8}R^{3}.\label{eq:Cubic_curvature_Lagrangian}
\end{align}
 Here, note that we have chosen $R^{pqrs}R_{p\phantom{t}r}^{\phantom{p}t\phantom{r}u}R_{qust}$
specifically instead of $R^{pqrs}R_{p\phantom{t}r}^{\phantom{p}t\phantom{r}u}R_{qtsu}$
which was given in \cite{Fulling}. Because it is not possible to
put the latter in the up-up-down-down form, $R^{pqrs}R_{p\phantom{t}r}^{\phantom{p}t\phantom{r}u}R_{qtsu}=R_{rs}^{pq}R_{pt}^{ru}R_{q\phantom{ts}u}^{\phantom{q}ts}$.
But, the two terms are related to each other as 
\begin{equation}
R^{pqrs}R_{p\phantom{t}r}^{\phantom{p}t\phantom{r}u}R_{qtsu}=R^{pqrs}R_{p\phantom{t}r}^{\phantom{p}t\phantom{r}u}R_{qust}+\frac{1}{4}R^{pqrs}R_{pq}^{\phantom{pq}tu}R_{rstu}.\label{eq:Relating_basis}
\end{equation}
 Leaving the details of the computation to the Appendix B, here we
summarize the final results. The equivalent quadratic action that
has the same free theory as (\ref{eq:Cubic_action}) becomes 
\begin{equation}
I=\int d^{D}x\sqrt{-g}\,\left[\frac{1}{\tilde{\kappa}}\left(R-2\tilde{\Lambda}_{0}\right)+\tilde{\alpha}R^{2}+\tilde{\beta}R_{ab}^{2}+\tilde{\gamma}\left(R_{abcd}^{2}-4R_{ab}^{2}+R^{2}\right)\right],\label{eq:Equiv_quad_act}
\end{equation}
 with parameters 
\begin{align}
\frac{1}{\tilde{\kappa}}\equiv & \frac{1}{\kappa}-\frac{12\Lambda^{2}}{\left(D-1\right)^{2}\left(D-2\right)^{2}}\nonumber \\
 & \times\left[\left(D-3\right)a_{1}+4a_{2}+2\left(D-1\right)\left(a_{3}+Da_{4}\right)+\left(D-1\right)^{2}\left(a_{5}+a_{6}+Da_{7}+D^{2}a_{8}\right)\right],\label{eq:keff}\\
\tilde{\Lambda}_{0}\equiv & \frac{\tilde{\kappa}}{\kappa}\Lambda_{0}+\frac{D\Lambda}{3\left(D-2\right)}\left(1-\frac{\tilde{\kappa}}{\kappa}\right),\label{eq:Leff}\\
\tilde{\alpha}\equiv & \alpha+\frac{2\Lambda}{\left(D-1\right)\left(D-2\right)}\nonumber \\
 & \times\left[3a_{1}-6a_{2}-\left(D^{2}-D-4\right)a_{4}+a_{5}+\left(D-1\right)\left(-a_{3}+2a_{7}+3Da_{8}\right)\right],\label{eq:aeff}\\
\tilde{\beta}\equiv & \beta+\frac{2\Lambda}{\left(D-1\right)\left(D-2\right)}\nonumber \\
 & \times\left[-9a_{1}+24a_{2}+4Da_{3}+\left(2D-3\right)a_{5}+\left(D-1\right)\left(4Da_{4}+3a_{6}+Da_{7}\right)\right],\label{eq:beff}\\
\tilde{\gamma}\equiv & \gamma+\frac{2\Lambda}{\left(D-1\right)\left(D-2\right)}\left[-3a_{1}+6a_{2}+\left(D-1\right)\left(a_{3}+Da_{4}\right)\right].\label{eq:geff}
\end{align}
 The equation that determines the effective cosmological constant
is 
\begin{equation}
\frac{\Lambda-\tilde{\Lambda}_{0}}{2\tilde{\kappa}}+\tilde{f}\Lambda^{2}=0,\qquad\qquad\tilde{f}\equiv\left(D\tilde{\alpha}+\tilde{\beta}\right)\frac{\left(D-4\right)}{\left(D-2\right)^{2}}+\tilde{\gamma}\frac{\left(D-3\right)\left(D-4\right)}{\left(D-1\right)\left(D-2\right)},\label{eq:Vacuum_of_eqa}
\end{equation}
 which, generically is a cubic order equation in $\Lambda$ once $\tilde{\kappa}$
and $\tilde{f}$ are explicitly written. To check the unitarity, one
also needs the equivalent effective Newton's constant 
\begin{equation}
\frac{1}{\tilde{\kappa}_{e}}\equiv\frac{1}{\tilde{\kappa}}+\frac{4\Lambda D}{D-2}\tilde{\alpha}+\frac{4\Lambda}{D-1}\tilde{\beta}+\frac{4\Lambda\left(D-3\right)\left(D-4\right)}{\left(D-1\right)\left(D-2\right)}\tilde{\gamma}.\label{eq:ee-Newton}
\end{equation}
 The equations from (\ref{eq:keff}) to (\ref{eq:ee-Newton}) can
be used to classify all the unitary cubic curvature theories. By choosing
the parameters of the equivalent quadratic theory, one can reduce
it to any one of those \textcolor{black}{four} unitary theories listed
in the introduction and studied in Sec. \ref{sec:Propagator}. For
example, if $\tilde{\alpha}=\tilde{\beta}=\tilde{\gamma}=0$, the
free theory of the cubic curvature gravity reduces to the cosmological
Einstein theory with a massless spin-2 degree of freedom. These three
conditions (which are linearly independent when $\alpha$, $\beta$,
and $\gamma$ do not vanish) reduce the 13 parameters of the cubic
theory to 10 parameters. The unitarity condition $\tilde{\kappa}>0$
gives a constraint on these 10 parameters yielding a wide range of
unitary cubic curvature theories. Let us note that as long as any
one of the lower order parameters $\frac{1}{\kappa}$, $\Lambda_{0}$,
$\alpha$, $\beta$, and $\gamma$ do not vanish, the cubic equation
(\ref{eq:Vacuum_of_eqa}) has at least \emph{one} real root, meaning
that the theory has a maximally symmetric vacuum with a nonzero curvature.
Similarly, the cubic curvature theory can be reduced to the remaining
four unitary theories, each time with only a couple of constraints
on the parameters. This result classifies all the unitary cubic curvature
theories in $D$ dimensional (A)dS spacetimes. Therefore, it could
also be used to check the unitarity of the existing cubic curvature
gravity theories found with different guiding principles. Below we
will study some examples such as the cubic curvature gravity obtained
as an effective theory in the bosonic string theory \cite{Tseytlin},
and the theories introduced in \cite{Oliva1,Oliva2} and \cite{MyersRobinson,MyersSinha1,MyersSinha2}.

\textcolor{black}{Using (\ref{eq:keff}-\ref{eq:ee-Newton}), one
can also find the cubic curvature extensions of the recently found
critical gravity \cite{LuPope,DeserLiu} which is given in Appendix
A.}

\subsection{The purely cubic theory}

In \cite{Oliva1,MyersRobinson}, the following cubic curvature theory
was introduced 
\begin{align}
I=a\int d^{5}x\sqrt{-g}\, & \biggl(-\frac{7}{6}R_{cd}^{ab}R_{bf}^{ce}R_{ae}^{df}-R_{ab}^{cd}R_{cd}^{be}R_{e}^{a}-\frac{1}{2}R^{pq}R^{rs}R_{prqs}\nonumber \\
 & +\frac{1}{3}R^{pq}R_{p}^{r}R_{qr}-\frac{1}{2}RR^{pq}R_{pq}+\frac{1}{12}R^{3}\biggr).\label{eq:quasi_topo}
\end{align}
 The trace of the field equations coming from this action is proportional
to the Lagrangian, and the action has just the first derivative of
$g^{rr}\left(r\right)$ in the spherically symmetric ansatz. Note
that there is no linear and quadratic terms in the curvature. One
does not expect this homogeneous (in curvature) theory to have an
(A)dS vacuum. This can easily be seen from the equivalent quadratic
Lagrangian which follows from (\ref{eq:Equiv_quad_act}) 
\begin{align}
f_{\text{quad-equal}}\left(R_{cd}^{ab}\right)= & -\frac{\Lambda^{2}}{12}\left(R-\frac{10\Lambda}{9}\right)+\frac{\Lambda}{12}\left(R_{abcd}^{2}-4R_{ab}^{2}+R^{2}\right).\label{eq:eqa_quasi-topo}
\end{align}
 Using (\ref{eq:Vacuum_of_eqa}), one obtains $\Lambda=0$ which says
that the equivalent quadratic action (\ref{eq:eqa_quasi-topo}) vanishes.
Therefore, there is no propagator in this theory and there are no
local degrees of freedom, unless one introduces linear and quadratic
or zeroth order terms in the curvature. (It is likely that these theories
have local degrees of freedom at the nonlinear level as in the case
of Chern-Simons theories which lack local degrees of freedom at the
quadratic level, but have nonzero degrees of freedom at the nonlinear
level \cite{Banados}.) In fact, for any \emph{purely} cubic curvature
theory, unless the trace of the field equations vanishes identically,
one does not have local degrees of freedom. {[}Taking the risk of
being too pedantic, let us note that this fact follows from the vanishing
of \emph{$O\left(h^{2}\right)$ }expansion of \emph{any} cubic or
higher order curvature invariant around flat space.{]}

The generalization of (\ref{eq:quasi_topo}) to $D$ dimensions was
given in \cite{MyersRobinson,Oliva2,MyersSinha2} in different forms.
Here, we choose the action given in \cite{MyersRobinson} 
\begin{align}
\mathcal{Z}_{D}=a\int d^{D}x\sqrt{-g}\, & \Biggl[R^{pqrs}R_{p\phantom{t}r}^{\phantom{p}t\phantom{r}u}R_{qtsu}+\frac{1}{\left(2D-3\right)\left(D-4\right)}\nonumber \\
 & \times\biggl(-3\left(D-2\right)R^{pq}R_{\phantom{rst}p}^{rst}R_{rstq}+\frac{3\left(3D-8\right)}{8}RR^{pqrs}R_{pqrs}\label{eq:MR_D}\\
 & +3DR^{pq}R^{rs}R_{prqs}+6\left(D-2\right)R^{pq}R_{p}^{r}R_{qr}-\frac{3\left(3D-4\right)}{2}RR^{pq}R_{pq}+\frac{3D}{8}R^{3}\Biggr)\Biggr].\nonumber 
\end{align}
 In fact, a second action was also given in \cite{MyersRobinson}
\begin{align}
\mathcal{Z}_{D}^{\prime}=a\int d^{D}x\sqrt{-g}\, & \Biggl[R^{pqrs}R_{pq}^{\phantom{pq}tu}R_{rstu}+\frac{1}{\left(2D-3\right)\left(D-4\right)}\biggl(-12\left(D^{2}-5D+5\right)R^{pq}R_{\phantom{rst}p}^{rst}R_{rstq}\nonumber \\
 & +\frac{3}{2}\left(D^{2}-4D+2\right)RR^{pqrs}R_{pqrs}+12\left(D-2\right)\left(D-3\right)R^{pq}R^{rs}R_{prqs}\\
 & +8\left(D-1\right)\left(D-3\right)R^{pq}R_{p}^{r}R_{qr}-6\left(D-2\right)^{2}RR^{pq}R_{pq}+\frac{1}{2}\left(D^{2}-4D+6\right)R^{3}\Biggr)\Biggr],\nonumber 
\end{align}
 which is related with the previous one as $\mathcal{Z}_{D}^{\prime}=2\mathcal{Z}_{D}+\frac{1}{4}\mathcal{X}_{6}$
where $\mathcal{X}_{6}$ is the six-dimensional Euler density 
\begin{align}
\mathcal{X}_{6}= & -8R^{pqrs}R_{p\phantom{t}r}^{\phantom{p}t\phantom{r}u}R_{qtsu}+4R^{pqrs}R_{pq}^{\phantom{pq}tu}R_{rstu}-24R^{pq}R_{\phantom{rst}p}^{rst}R_{rstq}\nonumber \\
 & +3RR^{pqrs}R_{pqrs}+24R^{pq}R^{rs}R_{prqs}+16R^{pq}R_{p}^{r}R_{qr}-12RR^{pq}R_{pq}+R^{3}.\label{eq:ED6}
\end{align}
 To find the equivalent quadratic action for $\mathcal{Z}_{D}$, we
first use (\ref{eq:Relating_basis}) to see the Lagrangian as a function
of $R_{cd}^{ab}$. Then, from (\ref{eq:keff}, \ref{eq:Leff}, \ref{eq:aeff},
\ref{eq:beff}, \ref{eq:geff}), one arrives at the cosmological Einstein-Gauss-Bonnet
theory 
\begin{align}
f_{\text{quad-equal}}\left(R_{cd}^{ab}\right)= & -\frac{3\left(D-3\right)\left(3D^{2}-15D+16\right)\Lambda^{2}}{2\left(2D-3\right)\left(D-1\right)^{2}\left(D-2\right)}a\left(R-\frac{2D\Lambda}{3\left(D-2\right)}\right)\label{eq:eqa_ZD}\\
 & +\frac{3\left(3D^{2}-15D+16\right)\Lambda}{4\left(D-1\right)\left(D-2\right)\left(2D-3\right)}a\left(R_{abcd}^{2}-4R_{ab}^{2}+R^{2}\right).\nonumber 
\end{align}
 Again, one finds $\Lambda=0$ from (\ref{eq:Vacuum_of_eqa}), and
the theory has no propagating degrees of freedom. The equivalent quadratic
action for $\mathcal{Z}_{D}^{\prime}$ is proportional to (\ref{eq:eqa_ZD})
with a proportionality constant $\frac{4\left(D^{3}-9D^{2}+26D-22\right)}{3D^{2}-15D+16}$.
Therefore, as expected in the purely cubic curvature theories, it,
too, has no local degrees of freedom. On the other hand, if we add
lower powers of curvature to $\mathcal{Z}_{D}$ or $\mathcal{Z}_{D}^{\prime}$,
breaking homogeneity, then we have theories with local degrees of
freedom. As an example, let us consider $\mathcal{Z}_{D}$ augmented
with all the possible lower order terms 
\begin{equation}
I=\int d^{D}x\,\sqrt{-g}\left[\frac{1}{\kappa}\left(R-2\Lambda_{0}\right)+\alpha R^{2}+\beta R_{ab}^{2}+\gamma\left(R_{abcd}^{2}-4R_{ab}^{2}+R^{2}\right)\right]+\mathcal{Z}_{D}.
\end{equation}
 The equivalent quadratic action has the following parameters 
\begin{align}
\frac{1}{\tilde{\kappa}} & \equiv\frac{1}{\kappa}-\frac{3\left(D-3\right)\left(3D^{2}-15D+16\right)\Lambda^{2}}{2\left(2D-3\right)\left(D-1\right)^{2}\left(D-2\right)}a,\qquad\tilde{\Lambda}_{0}\equiv\frac{\tilde{\kappa}}{\kappa}\Lambda_{0}+\frac{D\Lambda}{3\left(D-2\right)}\left(1-\frac{\tilde{\kappa}}{\kappa}\right),\nonumber \\
\tilde{\alpha} & \equiv\alpha,\qquad\tilde{\beta}\equiv\beta,\qquad\tilde{\gamma}\equiv\gamma+\frac{3\left(3D^{2}-15D+16\right)\Lambda}{4\left(D-1\right)\left(D-2\right)\left(2D-3\right)}a,
\end{align}
 where $\Lambda$ is determined by (\ref{eq:Vacuum_of_eqa}), but
the result is not particularly illuminating to depict. Let us give
a simple specific example. Consider $D=5$, $\alpha=\beta=\gamma=\Lambda_{0}=0$,
then the three roots are given as 
\begin{equation}
\Lambda=0,\qquad\Lambda_{\pm}=\pm3\sqrt{\frac{7}{\kappa a}}.
\end{equation}
 To have an (A)dS vacuum, $\kappa a$ should be positive. For $\Lambda_{\pm}$,
$\tilde{\kappa}=-\frac{\kappa}{8}$ and $\tilde{\kappa}_{e}=-\frac{\kappa}{2}$,
therefore, to have $\kappa_{e}>0$, we must choose $\kappa<0$ and
$a<0$. The theory is unitary for these values, but it has the wrong
sign bare Newton's constant. Of course, around $\Lambda=0$, the theory
is unitary for $\kappa>0$, but the cubic terms do not contribute
and the free theory is equivalent to the Einstein-Hilbert gravity.
With nonvanishing $\Lambda_{0}$, the picture changes, there are regions
where the theory is unitary for $\kappa>0$. Similar analysis can
be done in $D$ dimensions, but the punchline of this section is that
the equivalent quadratic actions of $\mathcal{Z}_{D}$ and $\mathcal{Z}_{D}^{\prime}$
are just the Einstein-Gauss-Bonnet theories.

\subsection{Cubic theory of Oliva--Ray and Myers--Sinha}

In \cite{Oliva2}, a cubic curvature theory was introduced using the
requirement that the trace of the field equations is a second order
equation of the metric whose cubic curvature action in our notation
reads 
\begin{equation}
I=\int d^{D}x\sqrt{-g}\,\left[\frac{1}{\kappa}\left(R-2\Lambda_{0}\right)+\gamma\left(R_{abcd}^{2}-4R_{ab}^{2}+R^{2}\right)+b_{1}W_{1}+b_{2}W_{2}+b_{3}\chi_{6}\right],\label{eq:MS}
\end{equation}
 where $W_{1,2}$ are the inequivalent (for $D>5$) contractions of
three Weyl tensors 
\begin{equation}
W_{1}=C^{abcd}C_{a\phantom{e}c}^{\phantom{a}e\phantom{c}f}C_{bedf}\qquad W_{2}=C^{abcd}C_{ab}^{\phantom{ab}ef}C_{cdef},
\end{equation}
 and $\chi_{6}$ was given in (\ref{eq:ED6}). The same theory was
introduced using AdS/CFT and the existence of a \emph{simple} holographic
$c$-function in \cite{MyersSinha2}. To find the equivalent quadratic
action of this theory and study the unitarity, there are two ways.
The first way is to express $W_{1}$ and $W_{2}$ in terms of the
tensor structures that appear in (\ref{eq:Cubic_curvature_Lagrangian})
and use (\ref{eq:Equiv_quad_act}). We have carried out this lengthy
computation, but since the result is the same as a simpler computation
that we shall lay out now, we will not present it here. The second,
relatively simple, way is to realize that $W_{1}$ and $W_{2}$ do
not contribute to the propagator in (A)dS or any conformally flat
space as a matter of fact, and therefore, from (\ref{eq:keff}, \ref{eq:Leff},
\ref{eq:aeff}, \ref{eq:beff}, \ref{eq:geff}) one can find the equivalent
quadratic action as 
\begin{align}
f_{\text{quad-equal}}\left(R_{cd}^{ab}\right)= & -\frac{2\Lambda_{0}}{\kappa}+\frac{8D\left(D-3\right)\left(D-4\right)\left(D-5\right)\Lambda^{3}}{\left(D-1\right)^{2}\left(D-2\right)^{2}}b_{3}\nonumber \\
 & +\left[\frac{1}{\kappa}-\frac{12\left(D-3\right)\left(D-4\right)\left(D-5\right)\Lambda^{2}}{\left(D-1\right)^{2}\left(D-2\right)}b_{3}\right]R\label{eq:eqa_ED6}\\
 & +\left[\gamma+\frac{6\left(D-4\right)\left(D-5\right)\Lambda}{\left(D-1\right)\left(D-2\right)}b_{3}\right]\left(R_{abcd}^{2}-4R_{ab}^{2}+R^{2}\right).\nonumber 
\end{align}
 Therefore, the free theory of (\ref{eq:MS}) boils down to the EGB
theory classified as the second unitary theory in the introduction.
(The coefficient of the Gauss-Bonnet term can also vanish for a tuned
value of $b_{3}$ yielding an interesting theory that we deal below.)
The effective cosmological constant satisfies 
\begin{equation}
\frac{\Lambda-\Lambda_{0}}{2\kappa}+\frac{\gamma\Lambda^{2}\left(D-3\right)\left(D-4\right)}{\left(D-1\right)\left(D-2\right)}+\frac{2b_{3}\Lambda^{3}\left(D-3\right)\left(D-4\right)\left(D-5\right)\left(D-6\right)}{\left(D-1\right)^{2}\left(D-2\right)^{2}}=0,\label{eq:vac_MS}
\end{equation}
 which has at least one real root for any value of the parameters
in generic $D>6$ dimensions. As expected for $D\le6$, Euler density
$\chi_{6}$ does not contribute and for $D\le4$ the Gauss-Bonnet
term does not contribute. The unitarity constraint 
\begin{equation}
\frac{1}{\tilde{\kappa}_{e}}=\frac{1}{\kappa}+\frac{4\Lambda\gamma\left(D-3\right)\left(D-4\right)}{\left(D-1\right)\left(D-2\right)}+\frac{12b_{3}\Lambda^{2}\left(D-3\right)\left(D-4\right)\left(D-5\right)\left(D-6\right)}{\left(D-1\right)^{2}\left(D-2\right)^{2}}>0,
\end{equation}
 which somewhat simplifies to 
\begin{equation}
\frac{1}{\tilde{\kappa}_{e}}=\frac{3\Lambda_{0}}{\kappa\Lambda}-\frac{2}{\kappa}-\frac{2\Lambda\gamma\left(D-3\right)\left(D-4\right)}{\left(D-1\right)\left(D-2\right)}>0,\label{eq:ket_MS}
\end{equation}
 restricts the parameters, but there is still a wide range of unitarity
region. Let us specifically consider $D=7$, and $\Lambda_{0}=\gamma=0$,
for which one gets from (\ref{eq:vac_MS}) 
\begin{equation}
\Lambda=0,\qquad\Lambda=\pm\frac{5}{2}\sqrt{-\frac{3}{2\kappa b_{3}}}.
\end{equation}
 The existence of a constant curvature vacuum requires $\kappa b_{3}<0$.
Assuming $\kappa>0$ to have the usual Einstein-Hilbert gravity in
IR, one should choose $b_{3}<0$. However, this is in conflict with
the UV unitarity, since $\tilde{\kappa}_{e}=-\frac{\kappa}{2}<0$.
Again, assuming nonzero $\Lambda_{0}$ or $\gamma$ changes the picture
and allows unitary regions which can be found from (\ref{eq:vac_MS})
and (\ref{eq:ket_MS}). Moreover, one can add $\alpha R^{2}$ and
$\beta R_{ab}^{2}$ to the action, and still have unitary regions.

Let us now consider the case for which (\ref{eq:MS}) reduces to the
pure Einstein-Hilbert gravity at the free-level. For this case, one
should tune the coefficient of $\chi_{6}$ (assuming $D>6$) 
\begin{equation}
b_{3}=-\frac{\gamma\left(D-1\right)\left(D-2\right)}{6\Lambda\left(D-4\right)\left(D-5\right)}.
\end{equation}
 Then, (\ref{eq:vac_MS}) has two solutions 
\begin{equation}
\Lambda_{\pm}=-\frac{3\left(D-1\right)\left(D-2\right)}{8\kappa\gamma\left(D-3\right)^{2}}\left(1\pm\sqrt{1+\frac{16\kappa\gamma\Lambda_{0}\left(D-3\right)^{2}}{3\left(D-1\right)\left(D-2\right)}}\right).
\end{equation}
 Further imposing the uniqueness of the vacuum $\Lambda_{+}=\Lambda_{-}$,
one has $\Lambda=2\Lambda_{0}$ and a tuned Gauss-Bonnet coefficient
\begin{equation}
\gamma=-\frac{3\left(D-1\right)\left(D-2\right)}{16\kappa\Lambda_{0}\left(D-3\right)^{2}}.
\end{equation}
 The theory has an equivalent effective Newton's constant 
\begin{equation}
\tilde{\kappa}_{e}=\tilde{\kappa}=\frac{4\kappa\left(D-3\right)}{\left(D-6\right)}>0.
\end{equation}
 Summing up the result: the following cubic action has a unique (A)dS
vacuum around which it has a unitary massless spin-2 mode 
\begin{align}
I=\int d^{D}x\sqrt{-g}\, & \Biggl[\frac{1}{\kappa}\left(R-2\Lambda_{0}\right)-\frac{3\left(D-1\right)\left(D-2\right)}{16\kappa\Lambda_{0}\left(D-3\right)^{2}}\left(R_{abcd}^{2}-4R_{ab}^{2}+R^{2}\right)\nonumber \\
 & +b_{1}W_{1}+b_{2}W_{2}+\frac{\left(D-1\right)^{2}\left(D-2\right)^{2}}{64\kappa\Lambda_{0}^{2}\left(D-3\right)^{2}\left(D-4\right)\left(D-5\right)}\chi_{6}\Biggr].\label{eq:MS-special}
\end{align}
 Furthermore, for $\Lambda_{0}<0$, the theory is also unitary in
the boundary of AdS, since the free theory is exactly like the cosmological
Einstein's gravity with modified parameters: 
\begin{equation}
I_{\text{quad-equal}}=\int d^{D}x\sqrt{-g}\left[\frac{D-6}{4\kappa\left(D-3\right)}\left(R-4\Lambda_{0}\right)\right].\label{eq:eqa_MS-special}
\end{equation}
 The fact that (\ref{eq:MS-special}) and (\ref{eq:eqa_MS-special})
have the same $O\left(h^{0}\right)$, $O\left(h\right)$ and $O\left(h^{2}\right)$
expansions is quite remarkable, and is a testament to the effectiveness
of the tools laid out in this paper in studying the spectra and the
vacua of higher curvature gravity theories.

\subsection{Cubic gravity generated by string theory}

In \cite{Tseytlin}, using the three and four point scattering amplitudes
of strings, it was shown that heterotic and type-II superstring theories
do not produce any $R^{3}$ interactions, but the bosonic string has
the following (in our basis) effective action 
\begin{equation}
I=\frac{1}{\kappa}\int d^{D}x\sqrt{-g}\,\left[R+\frac{\alpha^{\prime}}{4}\left(R_{abcd}^{2}-4R_{ab}^{2}+R^{2}\right)+\frac{\left(\alpha^{\prime}\right)^{2}}{24}\left(-2R^{pqrs}R_{p\phantom{t}r}^{\phantom{p}t\phantom{r}u}R_{qust}+R^{pqrs}R_{pq}^{\phantom{pq}tu}R_{rstu}\right)\right],\label{eq:Effective_string}
\end{equation}
 where $\alpha^{\prime}$ is the usual inverse string tension. In
this action, the coefficients of $R_{ab}^{2}$ and $R^{2}$ are not
unambiguously determined by the computation of \cite{Tseytlin}, but
chosen for their simplicity. In our analysis, we will keep this choice;
but, without much effort, one can of course consider these coefficients
to be arbitrary. Note also that there is no bare cosmological constant.
Let us now consider (\ref{eq:Effective_string}) to be our microscopic
theory, forgetting its string theory origin as an effective theory,
and study its spectrum and unitarity. Clearly, around flat space,
it describes a massless spin-2 unitary excitation. Around its assumed
(A)dS vacua, the equivalent quadratic action follows from (\ref{eq:Equiv_quad_act})
\begin{align}
I_{\text{quad-equal}}=\frac{1}{\kappa}\int d^{D}x\sqrt{-g}\, & \Biggl[-\frac{2\alpha^{\prime2}\Lambda^{3}D\left(D-5\right)}{3\left(D-1\right)^{2}\left(D-2\right)^{3}}+\left(1+\frac{\alpha^{\prime2}\Lambda^{2}\left(D-5\right)}{\left(D-1\right)^{2}\left(D-2\right)^{2}}\right)R\nonumber \\
 & +\frac{\alpha^{\prime}}{4}\left(R_{abcd}^{2}-4R_{ab}^{2}+R^{2}\right)+\frac{\alpha^{\prime2}\Lambda}{2\left(D-1\right)\left(D-2\right)}\left(2R_{abcd}^{2}-R_{ab}^{2}\right)\Biggr].\label{eq:eqa_eff_string}
\end{align}
 As before, $\Lambda$ can be determined from the vacuum equation
(\ref{eq:Vacuum_of_eqa}). Specifically, for the critical dimension
of the bosonic string $D=26$, we have $\Lambda=0$, $\alpha^{\prime}\Lambda\approx-2.37133$,
and $\alpha^{\prime}\Lambda\approx26025.2$, but because of the last
term in (\ref{eq:eqa_eff_string}), the theory is not unitary in (A)dS.
Even if one adds %
\footnote{ One might be worried about $\alpha^{\prime2}$ appearing at the quadratic
level, but $\Lambda$ is $O\left(\frac{1}{\alpha^{\prime}}\right)$
for $D=26$. In fact, one obtains $\Lambda=0$, $\alpha^{\prime}\Lambda\approx-2.41$,
and $\alpha^{\prime}\Lambda\approx-145.6$ (No dS vacuum!).%
} 
\begin{equation}
\frac{\alpha^{\prime2}\Lambda}{\kappa\left(D-1\right)\left(D-2\right)}\int d^{D}x\sqrt{-g}\left(R^{2}-\frac{7}{2}R_{ab}^{2}\right)
\end{equation}
 to the cubic action to turn the free part of the theory to the EGB
theory with only a massless spin-2 excitation, the effective Newton's
constant $\tilde{\kappa}_{e}$ becomes negative, meaning, one has
a massless spin-2 ghost. Therefore, the theory is not unitary.

\section{Conclusion}

We have systematically constructed all the perturbatively unitary
cubic curvature gravity theories in $D$-dimensional (A)dS spacetimes.
Our construction is based on finding an equivalent quadratic action
that has the same vacua and the propagator as the original cubic theory.
This equivalent quadratic action follows from a Taylor series expansion
of the cubic action in powers of curvature up to second order. Using
our general result, which was given in (\ref{eq:Equiv_quad_act}),
we have studied the unitarity of the cubic curvature theories that
appeared in the literature recently. These are the theories studied
in \cite{Oliva1,MyersRobinson,Oliva2,MyersPaulosSinha,MyersSinha1,MyersSinha2}
that satisfy quite remarkable properties such as having second order
equations and admitting simple $c$-functions. We have also studied
the bosonic string generated cubic curvature gravity of \cite{Tseytlin}
and found that it is nonunitary in (A)dS. We have also given unique
vacuum cubic curvature extensions of the quadratic critical gravity
theories. The tools we have presented here are very powerful in studying
the free spectrum and the vacua of higher derivative gravity theories
in (A)dS, and can be extended to any higher order. In fact, we have
employed these tools in the infinite order Born-Infeld gravity theories
\cite{Gullu-BIUnitarity}, and the $D$-dimensional Lovelock gravities
\cite{Gullu-Lovelock}.

\section{Acknowledgments}

This work is supported by the T{Ü}B\.{I}TAK Grant No. 110T339, and
METU Grant BAP-07-02-2010-00-02. We would like to thank Arkady Tseytlin
for useful discussions on the cubic gravity generated by string theory.
We would like to thank Sourya Ray for commenting on \cite{Oliva1}.

\appendix

\section{Scattering Amplitude and Cubic Curvature Extension of Critical Gravity}

\textcolor{black}{Critical gravity \cite{LuPope,DeserLiu} is defined
by the action 
\begin{equation}
I=\int d^{D}x\,\sqrt{-g}\left[\frac{1}{\kappa}\left(R-2\Lambda_{0}\right)+\gamma C_{abcd}^{2}\right],\label{weylaction}
\end{equation}
 where the bare cosmological constant should be tuned to the value
$\Lambda_{0}=\frac{\left(D-1\right)\left(D-2\right)}{8\kappa\gamma\left(D-3\right)}=\Lambda$,
and $C_{abcd}$ is the Weyl tensor. The spectrum of the theory involves
only massless spin-2 degree of freedom. The effective Newton's constant
is given as $\kappa_{e}=\kappa\left(1-\frac{D}{2}\right)$ which has
the opposite sign compared to the bare one. As we will see, the propagator
of this theory has a double pole at the zero on-shell mass.}

\subsection{Scattering in the critical gravity and its unitarity}

Scattering in critical gravity theory is a somewhat subtle issue,
this is because critical theory was defined for $T_{ab}=0$, and the
criticality strictly depends on $R_{L}=-\frac{2\Lambda}{D-2}h=0$.
If one wants to extend the critical theory to nonzero $T_{ab}$, then,
apparently, one should restrict it to the case when the trace of the
energy-momentum tensor is zero, $T=0$. For example, in four dimensions,
Maxwell theory satisfies this condition and provides an example of
an extension of the critical theory with sources. First of all, in
the parametrization of (\ref{eq:General_quad_curv}), the critical
theory corresponds to \cite{DeserLiu} 
\begin{equation}
\beta=-\frac{4\alpha\left(D-1\right)}{D},\qquad\Lambda=\Lambda_{0}=-\frac{D}{8\kappa\alpha},\qquad\alpha=-\frac{\gamma D\left(D-3\right)}{\left(D-1\right)\left(D-2\right)}.
\end{equation}
 Then, for $T=0$, the scattering amplitude reads 
\begin{equation}
A=2T_{ab}^{\prime}\left[\left(\beta\bar{\square}+\frac{1}{\kappa_{e}}\right)\left(\triangle_{L}^{(2)}-\frac{4\Lambda}{D-2}\right)\right]^{-1}T^{ab}.
\end{equation}
 The Lichnerowicz operator $\triangle_{L}^{(2)}$ acting on a \emph{traceless
}rank-2 tensor $\varphi_{ab}$ yields 
\begin{equation}
\triangle_{L}^{(2)}\varphi_{ab}=\left(-\bar{\square}+\frac{4\Lambda D}{\left(D-1\right)\left(D-2\right)}\right)\varphi_{ab},
\end{equation}
 and plugging the critical value of $\beta$ in the form $\beta=-\frac{\left(D-1\right)\left(D-2\right)}{4\Lambda\kappa_{e}}$,
one has the double pole structure 
\begin{equation}
A=-\frac{2}{\beta}T_{ab}^{\prime}\left[\left(\bar{\square}-\frac{4\Lambda}{\left(D-1\right)\left(D-2\right)}\right)\left(\bar{\square}-\frac{4\Lambda}{\left(D-1\right)\left(D-2\right)}\right)\right]^{-1}T^{ab}.\label{eq:Amp_critic}
\end{equation}
 \textcolor{black}{According to the old nomenclature, we have a degenerate
Pais-Uhlenbeck oscillator; therefore, the two poles cannot be separated.
With this observation, it is suggestive to think that the theory is
free from the ghost, from which all the higher curvature theories
(in $D>3$) with an $R_{ab}^{2}$ term suffer, at the tree-level as
long as $\beta>0$ (for AdS, this requires $\kappa<0$, and for dS
$\kappa>0$). However, in \cite{PorratiRoberts}, it is shown that
logarithmic modes mix, in the Hilbert space, with the homogeneous
(Einstein) modes; and if one tries to restrict the physical space
to homogeneous modes only, the physical space will only involve vacuum
state.}

Let us now consider the $T\ne0$ case in the critical gravity. In
this case, $h$ cannot be zero, instead it is algebraically determined
by the trace of the source 
\begin{equation}
h=\frac{\kappa T}{\Lambda},
\end{equation}
 which comes from (\ref{eq:Linearized_eom}). The degrees of freedom
change and the scattering amplitude is no longer given as (\ref{eq:Amp_critic}),
instead one has 
\begin{align}
A= & \frac{2}{\beta}T_{ab}^{\prime}\left[\left(\bar{\square}-\frac{4\Lambda}{\left(D-1\right)\left(D-2\right)}\right)\left(\triangle_{L}^{(2)}-\frac{4\Lambda}{D-2}\right)\right]^{-1}T^{ab}\nonumber \\
 & +\frac{2}{\beta\left(D-2\right)}T^{\prime}\left[\left(\bar{\square}-\frac{4\Lambda}{\left(D-1\right)\left(D-2\right)}\right)\left(\bar{\square}+\frac{4\Lambda}{D-2}\right)\right]^{-1}T\\
 & -\frac{1}{\Lambda\beta\left(D-2\right)}T^{\prime}\left(\bar{\square}-\frac{4\Lambda}{\left(D-1\right)\left(D-2\right)}\right)^{-1}T\nonumber \\
 & +\frac{4D}{\beta\left(D-1\right)\left(D-2\right)^{2}}T^{\prime}\left[\left(\bar{\square}-\frac{4\Lambda}{\left(D-1\right)\left(D-2\right)}\right)\left(\bar{\square}+\frac{2\Lambda D}{\left(D-1\right)\left(D-2\right)}\right)\right]^{-1}T.\nonumber 
\end{align}
 To see more explicitly the change in the degrees of freedom, let
us compute the second order action in $h_{ab}$. The computation can
be simplified by directly computing the linearized field equations
\begin{equation}
\mathcal{E}_{ab}\equiv\beta\left(\bar{\square}-\frac{4\Lambda}{\left(D-1\right)\left(D-2\right)}\right)\mathcal{G}_{ab}^{L}+\frac{\beta\left(D-2\right)}{2\left(D-1\right)}\left(\bar{g}_{ab}\bar{\square}-\bar{\nabla}_{a}\bar{\nabla}_{b}-\frac{2\Lambda}{D-2}\bar{g}_{ab}\right)R^{L},
\end{equation}
 and by integrating with the help of a reverse calculus of variations:
$-\frac{1}{2}\int d^{D}x\,\sqrt{-\bar{g}}h^{ab}\mathcal{E}_{ab}$
where the overall factor comes from the self-adjointness of the involved
operators. First, let us choose the gauge $\bar{\nabla}^{a}h_{ab}=\bar{\nabla}_{b}h$
as in \cite{LuPope}, then $R^{L}=-\frac{2\Lambda}{D-2}h$. In this
gauge, one has the linearized Ricci tensor 
\begin{equation}
R_{ab}^{L}=\frac{1}{2}\left(\bar{\nabla}_{a}\bar{\nabla}_{b}h-\bar{\square}h_{ab}+\frac{4D\Lambda}{\left(D-1\right)\left(D-2\right)}h_{ab}-\frac{4\Lambda}{\left(D-1\right)\left(D-2\right)}\bar{g}_{ab}h\right),
\end{equation}
 and the linearized Einstein tensor 
\begin{equation}
\mathcal{G}_{ab}^{L}=\frac{1}{2}\left(\bar{\nabla}_{a}\bar{\nabla}_{b}h-\bar{\square}h_{ab}\right)+\frac{2\Lambda}{\left(D-1\right)\left(D-2\right)}h_{ab}+\frac{\Lambda\left(D-3\right)}{\left(D-1\right)\left(D-2\right)}\bar{g}_{ab}h.
\end{equation}
 After using the gauge condition, one ends up with 
\begin{align}
T_{ab}= & -\frac{\beta}{2}\left(\bar{\square}-\frac{4\Lambda}{\left(D-1\right)\left(D-2\right)}\right)^{2}h_{ab}\nonumber \\
 & +\frac{\beta}{2}\bar{\square}\bar{\nabla}_{a}\bar{\nabla}_{b}h+\frac{\beta\Lambda\left(D-4\right)}{\left(D-1\right)\left(D-2\right)}\bar{\nabla}_{a}\bar{\nabla}_{b}h\\
 & -\frac{\beta\Lambda}{\left(D-1\right)\left(D-2\right)}\left[\bar{\square}-\frac{2\Lambda\left(D^{2}-5D+8\right)}{\left(D-1\right)\left(D-2\right)}\right]\bar{g}_{ab}h.\nonumber 
\end{align}
 After integration by parts, the second order action in this source-coupled
critical theory becomes 
\begin{align}
I_{O\left(h^{2}\right)}=-\frac{1}{2}\int d^{D}x\,\sqrt{-\bar{g}}\Biggl\{ & -\frac{\beta}{2}h^{ab}\left(\bar{\square}-\frac{4\Lambda}{\left(D-1\right)\left(D-2\right)}\right)^{2}h_{ab}\nonumber \\
 & +\frac{\beta}{2}h\bar{\Box}^{2}h+\frac{\beta\Lambda\left(3D-7\right)}{\left(D-1\right)\left(D-2\right)}h\bar{\square}h+\frac{2\Lambda^{2}\beta\left(D^{2}-5D+8\right)}{\left(D-1\right)^{2}\left(D-2\right)^{2}}h^{2}\Biggr\},
\end{align}
 which clearly shows that when $h\ne0$, new degrees of freedom arise.
To actually see these degrees of freedom and their masses, one has
to do an \emph{orthogonal} decomposition of the $h_{ab}$ tensor as
in (\ref{eq:Decomposition}), and further decompose $h_{ab}^{TT}$
into {}``spatial'' tensor, {}``spatial'' vector, and {}``scalar''
parts in a suitable form of the (A)dS metric. This computation is
not needed for the purpose of this work. (For $D=3$, the computation
was carried out for generic quadratic theory in \cite{Gullu-Canonical}.)

\subsection{Cubic extensions of critical gravity}

\textcolor{black}{Using the equivalent quadratic action for cubic
curvature theories given in (\ref{eq:Equiv_quad_act}-\ref{eq:geff}),
let us discuss the cubic curvature extensions of the critical theory
found in \cite{LuPope,DeserLiu}.} To get a critical cubic curvature
gravity in (A)dS, the parameters of the equivalent quadratic action
should satisfy the following conditions. The first condition is 
\begin{equation}
\tilde{\beta}=-\frac{4\tilde{\alpha}\left(D-1\right)}{D},\label{eq:Remove_scalar_mode}
\end{equation}
 which removes the massive spin-0 mode, as long as 
\begin{equation}
\frac{1}{\tilde{\kappa}}+4\tilde{f}\Lambda\ne0,\label{eq:Exception}
\end{equation}
 otherwise the trace of the linearized field equations is automatically
zero, and therefore, $R_{L}$ is left undetermined. As a second condition,
the mass of the spin-2 excitation is set to zero: 
\begin{equation}
\frac{1}{\tilde{\kappa}_{e}}+\frac{4\Lambda\tilde{\beta}}{\left(D-1\right)\left(D-2\right)}=0,\label{eq:critical}
\end{equation}
 which upon use of (\ref{eq:Remove_scalar_mode}) yields 
\begin{equation}
0=\frac{1}{\tilde{\kappa}}+4\Lambda\tilde{f}+\frac{8\Lambda\tilde{\alpha}}{D}.\label{eq:critical_with_no_s0}
\end{equation}
 This second form makes the condition more transparent. In addition
to these the vacuum equation has to be satisfied

\begin{equation}
\frac{\Lambda-\tilde{\Lambda}_{0}}{2\tilde{\kappa}}+\tilde{f}\Lambda^{2}=0,\qquad\qquad\tilde{f}=\left[\left(D-1\right)\left(D-2\right)\tilde{\alpha}+D\left(D-3\right)\tilde{\gamma}\right]\frac{\left(D-4\right)}{D\left(D-1\right)\left(D-2\right)}.\label{eq:vacuum_of_eqa_ag}
\end{equation}
 In the case of the quadratic critical gravity, for which $\tilde{f}=f$
\emph{etc.}, (\ref{eq:critical_with_no_s0}) defines the unique critical
vacuum; therefore, one of the two roots of the second order vacuum
equation is always noncritical. This could be remedied by imposing
a unique critical vacuum which can be done in two ways: either one
has coalescing roots or has $f=0$. The first way contradicts (\ref{eq:critical_with_no_s0}),
but the second way leads to the unique vacuum critical theory

\begin{equation}
\beta=-\frac{4\alpha\left(D-1\right)}{D},\qquad\Lambda=\Lambda_{0}=-\frac{D}{8\kappa\alpha},\qquad\alpha=-\frac{\gamma D\left(D-3\right)}{\left(D-1\right)\left(D-2\right)}.\label{eq:Criticality_conds}
\end{equation}
 The cubic curvature extension is more subtle even though its equivalent
parameters satisfy the same equations (\ref{eq:Remove_scalar_mode}),
(\ref{eq:Exception}), (\ref{eq:critical_with_no_s0}), and (\ref{eq:vacuum_of_eqa_ag}).
The subtlety arises from the fact that tilde variables depend on $\Lambda$.
Therefore, (\ref{eq:critical_with_no_s0}) is a second order equation
in $\Lambda$ and apparently there are two critical vacua. Moreover,
(\ref{eq:vacuum_of_eqa_ag}) is a cubic equation in $\Lambda$ and
those two critical vacua should be roots of this equation. But, the
third root will always be noncritical (assuming the first two are
real). Therefore, just like in the case of the quadratic critical
theory, one should impose the unique vacuum condition. This again
can be done in two different ways: either one has coalescing roots
or has a linear equation. Again the first way is in contradiction
with (\ref{eq:critical_with_no_s0}). The second way leads to $\Lambda=\tilde{\Lambda}_{0}$=$\Lambda_{0}$
and $\tilde{\kappa}=\kappa$. As in the quadratic critical gravity,
$\Lambda_{0}$ is also tuned in terms of the other parameters. But,
in this cubic critical gravity case, it satisfies a quadratic equation
\begin{equation}
\frac{1}{\kappa}+\frac{8\Lambda_{0}\tilde{\alpha}}{D}=0,
\end{equation}
 where $\tilde{\alpha}$ is given in (\ref{eq:aeff}) (but, of course,
two of $a_{i}$'s are eliminated in terms of the others). Therefore,
for given parameters, say $\gamma$, $a_{1}$, $a_{2}$, $a_{3}$,
$a_{4}$, $a_{5}$, $a_{6}$, there are \emph{two} unique vacuum critical
cubic curvature theories. For completeness, let us write the action
\begin{align}
I=\int d^{D}x\sqrt{-g}\, & \Biggl[\frac{1}{\kappa}\left(R-2\Lambda_{0}\right)+\gamma C_{abcd}^{2}\nonumber \\
 & +\biggl(a_{1}R^{pqrs}R_{p\phantom{t}r}^{\phantom{p}t\phantom{r}u}R_{qust}+a_{2}R^{pqrs}R_{pq}^{\phantom{pq}tu}R_{rstu}+a_{3}R^{pq}R_{\phantom{rst}p}^{rst}R_{rstq}\label{eq:Critic_cubic}\\
 & \phantom{+}+a_{4}RR^{pqrs}R_{pqrs}+a_{5}R^{pq}R^{rs}R_{prqs}+a_{6}R^{pq}R_{p}^{r}R_{qr}+a_{7}RR^{pq}R_{pq}+a_{8}R^{3}\Biggr)\Biggr],\nonumber 
\end{align}
 with two of the parameters fixed as 
\begin{align}
a_{7}=\frac{1}{D\left(D-1\right)\left(D-2\right)} & \Biggl[-3\left(D-6\right)a_{1}-24a_{2}-4\left(2D-3\right)a_{3}\nonumber \\
 & -4D\left(D-1\right)a_{4}-\left(2D-3\right)\left(D-2\right)a_{5}-3\left(D-1\right)\left(D-2\right)a_{6}\Biggr],
\end{align}
 
\begin{align}
a_{8}=\frac{1}{D^{2}\left(D-1\right)^{2}\left(D-2\right)} & \Biggl[2\left(D^{2}-8D+6\right)a_{1}+4\left(5D-4\right)a_{2}+2\left(3D-4\right)\left(D-1\right)a_{3}\nonumber \\
 & +2D^{2}\left(D-1\right)a_{4}+\left(D-1\right)\left(D-2\right)^{2}a_{5}+2\left(D-1\right)^{2}\left(D-2\right)a_{6}\Biggr],
\end{align}
 and $\Lambda_{0}$ is a root of the following equation 
\begin{equation}
\frac{16\kappa D\left(D-3\right)\left[-3a_{1}+6a_{2}+\left(D-1\right)\left(a_{3}+Da_{4}\right)\right]}{\left(D-1\right)^{2}\left(D-2\right)^{2}}\Lambda_{0}^{2}+\frac{8\kappa\gamma D\left(D-3\right)}{\left(D-1\right)\left(D-2\right)}\Lambda_{0}-D=0,
\end{equation}
 which gives a further restriction on the parameters coming from the
reality of $\Lambda_{0}$. It is interesting to see that one can have
$\gamma=0$, and generate a Einstein plus cubic curvature theory whose
free theory is equivalent to the critical gravity, or one can have
$\gamma\ne0$ and extend the critical quadratic gravity to critical
cubic gravity.

\section{Calculation of Equivalent Quadratic Lagrangian Densities}

To find the equivalent quadratic Lagrangian of a generic higher curvature
gravity theory of the form $\mathcal{L}\equiv\sqrt{-g}F\left(R_{cd}^{ab}\right)$,
one needs the following expansion 
\begin{align}
f_{\text{quad-equal}}\left(R_{cd}^{ab}\right)= & F\left(\bar{R}_{cd}^{ab}\right)+\left[\frac{\partial F}{\partial R_{ln}^{km}}\right]_{\bar{R}_{ln}^{km}}\left(R_{ln}^{km}-\bar{R}_{ln}^{km}\right)\nonumber \\
 & +\frac{1}{2}\left[\frac{\partial^{2}F}{\partial R_{xz}^{wy}\partial R_{ln}^{km}}\right]_{\bar{R}_{ln}^{km}}\left(R_{ln}^{km}-\bar{R}_{ln}^{km}\right)\left(R_{xz}^{wy}-\bar{R}_{xz}^{wy}\right).\label{eq:eqa1}
\end{align}
 Sometimes, computationally, it is easier to consider the Ricci tensor
and the scalar as independent variables, and do the following expansion
instead 
\begin{align}
f_{\text{quad-equal}}\left(R,R_{b}^{a},R_{cd}^{ab}\right)= & F\left(\bar{R},\bar{R}_{b}^{a},\bar{R}_{cd}^{ab}\right)+\left[\frac{\partial F}{\partial R}\right]_{\bar{R}}\left(R-\bar{R}\right)+\left[\frac{\partial F}{\partial R_{j}^{i}}\right]_{\bar{R}}\left(R_{j}^{i}-\bar{R}_{j}^{i}\right)\nonumber \\
 & +\left[\frac{\partial F}{\partial R_{ln}^{km}}\right]_{\bar{R}}\left(R_{ln}^{km}-\bar{R}_{ln}^{km}\right)\nonumber \\
 & +\frac{1}{2}\left[\frac{\partial^{2}F}{\partial R^{2}}\right]_{\bar{R}}\left(R-\bar{R}\right)^{2}+\frac{1}{2}\left[\frac{\partial^{2}F}{\partial R_{j}^{i}\partial R_{l}^{k}}\right]_{\bar{R}}\left(R_{j}^{i}-\bar{R}_{j}^{i}\right)\left(R_{l}^{k}-\bar{R}_{l}^{k}\right)\nonumber \\
 & +\frac{1}{2}\left[\frac{\partial^{2}F}{\partial R_{xz}^{wy}\partial R_{ln}^{km}}\right]_{\bar{R}}\left(R_{ln}^{km}-\bar{R}_{ln}^{km}\right)\left(R_{xz}^{wy}-\bar{R}_{xz}^{wy}\right)\label{eq:eqa2}\\
 & +\left[\frac{\partial F}{\partial R\partial R_{j}^{i}}\right]_{\bar{R}}\left(R-\bar{R}\right)\left(R_{j}^{i}-\bar{R}_{j}^{i}\right)\nonumber \\
 & +\left[\frac{\partial F}{\partial R\partial R_{ln}^{km}}\right]_{\bar{R}}\left(R-\bar{R}\right)\left(R_{ln}^{km}-\bar{R}_{ln}^{km}\right)\nonumber \\
 & +\left[\frac{\partial F}{\partial R_{j}^{i}\partial R_{ln}^{km}}\right]_{\bar{R}}\left(R_{j}^{i}-\bar{R}_{j}^{i}\right)\left(R_{ln}^{km}-\bar{R}_{ln}^{km}\right).\nonumber 
\end{align}
 Of course (\ref{eq:eqa1}) and (\ref{eq:eqa2}) give the same results,
but we will use the latter below. Let us consider all the 8 cubic
curvature terms in (\ref{eq:Cubic_action}).

\subsection{Analysis of the three terms that do not involve the Riemann tensor}

Let us first find the equivalent quadratic Lagrangian for the simplest
term $F\left(R\right)\equiv R^{3}$ which reads 
\begin{equation}
f_{\text{quad-equal}}\left(R\right)=\frac{8D^{3}\Lambda^{3}}{\left(D-2\right)^{3}}-\frac{12D^{2}\Lambda^{2}}{\left(D-2\right)^{2}}R+\frac{6D\Lambda}{D-2}R^{2}.
\end{equation}
 The term constructed from the Ricci tensor alone $F\left(R_{b}^{a}\right)\equiv R_{q}^{p}R_{p}^{r}R_{r}^{q}$
has the following equivalent quadratic Lagrangian: 
\begin{equation}
f_{\text{quad-equal}}\left(R_{b}^{a}\right)=\frac{8D\Lambda^{3}}{\left(D-2\right)^{3}}-\frac{12\Lambda^{2}}{\left(D-2\right)^{2}}R+\frac{6\Lambda}{D-2}R_{ab}^{2}.
\end{equation}
 Finally, let us analyze the term involving the curvature scalar and
the Ricci tensor, that is $F\left(R,R_{b}^{a}\right)\equiv RR_{q}^{p}R_{p}^{q}$:
\begin{align}
f_{\text{quad-equal}}\left(R,R_{b}^{a}\right)= & \frac{8D^{2}\Lambda^{3}}{\left(D-2\right)^{3}}-\frac{12D\Lambda^{2}}{\left(D-2\right)^{2}}R+\frac{4\Lambda}{D-2}R^{2}+\frac{2D\Lambda}{D-2}R_{ab}^{2}.
\end{align}

\subsection{Separate analysis of the five terms involving Riemann tensor}

\subsubsection{The term $R_{rs}^{pq}R_{pt}^{ru}R_{qu}^{st}$:}

Let $F\left(R_{cd}^{ab}\right)\equiv R_{rs}^{pq}R_{pt}^{ru}R_{qu}^{st}$.
Then, the zeroth order part reads 
\begin{equation}
F\left(\bar{R}_{cd}^{ab}\right)=\bar{R}_{rs}^{pq}\bar{R}_{pt}^{ru}\bar{R}_{qu}^{st}=\frac{8D\left(D-3\right)\Lambda^{3}}{\left(D-1\right)^{2}\left(D-2\right)^{3}},
\end{equation}
 and the first order part reads 
\begin{equation}
\left[\frac{\partial F}{\partial R_{ln}^{km}}\right]_{\bar{R}}=3\bar{R}_{rk}^{pl}\bar{R}_{pm}^{rn}=\frac{12\Lambda^{2}}{\left(D-1\right)^{2}\left(D-2\right)^{2}}\left[\left(D-2\right)\delta_{k}^{l}\delta_{m}^{n}+\delta_{m}^{l}\delta_{k}^{n}\right].
\end{equation}
 In calculating the first order part, it is possible to take the derivatives
with respect to the Riemann tensor such that the symmetries of the
Riemann tensor are explicit. However, there is no need for such a
care, because these symmetries will be taken into account upon multiplication
with the difference terms $\left(R_{ln}^{km}-\bar{R}_{ln}^{km}\right)$
and $\left(R_{ln}^{km}-\bar{R}_{ln}^{km}\right)\left(R_{xz}^{wy}-\bar{R}_{xz}^{wy}\right)$
which satisfy the required symmetries, and the parts that do not obey
the symmetries give zero contribution. The second order part reads
\begin{align}
\left[\frac{\partial F}{\partial R_{xz}^{wy}\partial R_{ln}^{km}}\right]_{\bar{R}} & =3\left(\delta_{y}^{l}\delta_{k}^{z}\bar{R}_{wm}^{xn}+\bar{R}_{wk}^{xl}\delta_{y}^{n}\delta_{m}^{z}\right)\nonumber \\
 & =\frac{6\Lambda}{\left(D-1\right)\left(D-2\right)}\left(\delta_{y}^{l}\delta_{k}^{z}\delta_{w}^{x}\delta_{m}^{n}-\delta_{y}^{l}\delta_{k}^{z}\delta_{m}^{x}\delta_{w}^{n}+\delta_{w}^{x}\delta_{k}^{l}\delta_{y}^{n}\delta_{m}^{z}-\delta_{k}^{x}\delta_{w}^{l}\delta_{y}^{n}\delta_{m}^{z}\right).
\end{align}
 Therefore, the equivalent quadratic action for $F\left(R_{cd}^{ab}\right)\equiv R_{rs}^{pq}R_{pt}^{ru}R_{qu}^{st}$
becomes 
\begin{align}
f_{\text{quad-equal}}\left(R_{cd}^{ab}\right)= & \frac{8D\left(D-3\right)\Lambda^{3}}{\left(D-1\right)^{2}\left(D-2\right)^{3}}-\frac{12\left(D-3\right)\Lambda^{2}}{\left(D-1\right)^{2}\left(D-2\right)^{2}}R\nonumber \\
 & -\frac{6\Lambda}{\left(D-1\right)\left(D-2\right)}R_{abcd}^{2}+\frac{6\Lambda}{\left(D-1\right)\left(D-2\right)}R_{ab}^{2}.
\end{align}

\subsubsection{The term $R_{rs}^{pq}R_{pq}^{tu}R_{tu}^{rs}$:}

Let $F\left(R_{cd}^{ab}\right)\equiv R_{rs}^{pq}R_{pq}^{tu}R_{tu}^{rs}$.
Then, the zeroth order, the first order and the second order parts
read 
\begin{equation}
F\left(\bar{R}_{cd}^{ab}\right)=\bar{R}_{rs}^{pq}\bar{R}_{pq}^{tu}\bar{R}_{tu}^{rs}=\frac{32D\Lambda^{3}}{\left(D-1\right)^{2}\left(D-2\right)^{3}},
\end{equation}
 
\begin{equation}
\left[\frac{\partial F}{\partial R_{ln}^{km}}\right]_{\bar{R}}=3\bar{R}_{km}^{tu}\bar{R}_{tu}^{ln}=\frac{24\Lambda^{2}}{\left(D-1\right)^{2}\left(D-2\right)^{2}}\left(\delta_{k}^{l}\delta_{m}^{n}-\delta_{k}^{n}\delta_{m}^{l}\right),
\end{equation}
 
\begin{align}
\left[\frac{\partial F}{\partial R_{xz}^{wy}\partial R_{ln}^{km}}\right]_{\bar{R}} & =3\left(\delta_{k}^{x}\delta_{m}^{z}\bar{R}_{wy}^{ln}+\delta_{w}^{l}\delta_{y}^{n}\bar{R}_{km}^{xz}\right)\nonumber \\
 & =\frac{6\Lambda}{\left(D-1\right)\left(D-2\right)}\left(\delta_{k}^{x}\delta_{m}^{z}\delta_{w}^{l}\delta_{y}^{n}-\delta_{k}^{x}\delta_{m}^{z}\delta_{y}^{l}\delta_{w}^{n}+\delta_{w}^{l}\delta_{y}^{n}\delta_{k}^{x}\delta_{m}^{z}-\delta_{w}^{l}\delta_{y}^{n}\delta_{m}^{x}\delta_{k}^{z}\right),
\end{align}
 yielding an equivalent quadratic action 
\begin{equation}
f_{\text{quad-equal}}\left(R_{cd}^{ab}\right)=\frac{32D\Lambda^{3}}{\left(D-1\right)^{2}\left(D-2\right)^{3}}-\frac{48\Lambda^{2}}{\left(D-1\right)^{2}\left(D-2\right)^{2}}R+\frac{12\Lambda}{\left(D-1\right)\left(D-2\right)}R_{abcd}^{2}.
\end{equation}

\subsubsection{The term $R_{q}^{p}R_{tp}^{rs}R_{rs}^{tq}$:}

Let $F\left(R_{b}^{a},R_{cd}^{ab}\right)\equiv R_{q}^{p}R_{tp}^{rs}R_{rs}^{tq}$.
Then, the relevant derivatives are found as 
\begin{equation}
F\left(\bar{R}_{b}^{a},\bar{R}_{cd}^{ab}\right)=\bar{R}_{q}^{p}\bar{R}_{tp}^{rs}\bar{R}_{rs}^{tq}=\frac{16D\Lambda^{3}}{\left(D-1\right)\left(D-2\right)^{3}},
\end{equation}
 
\begin{equation}
\left[\frac{\partial F}{\partial R_{j}^{i}}\right]_{\bar{R}}=\bar{R}_{ti}^{rs}\bar{R}_{rs}^{tj}=\frac{8\Lambda^{2}}{\left(D-1\right)\left(D-2\right)^{2}}\delta_{i}^{j},
\end{equation}
 
\begin{equation}
\left[\frac{\partial F}{\partial R_{ln}^{km}}\right]_{\bar{R}}=\bar{R}_{q}^{n}\bar{R}_{km}^{lq}+\bar{R}_{m}^{p}\bar{R}_{kp}^{ln}=\frac{8\Lambda^{2}}{\left(D-1\right)\left(D-2\right)^{2}}\left(\delta_{k}^{l}\delta_{m}^{n}-\delta_{m}^{l}\delta_{k}^{n}\right),
\end{equation}
 
\begin{equation}
\left[\frac{\partial^{2}F}{\partial R_{xz}^{wy}\partial R_{ln}^{km}}\right]_{\bar{R}}=\bar{R}_{y}^{n}\delta_{w}^{l}\delta_{k}^{x}\delta_{m}^{z}+\bar{R}_{m}^{z}\delta_{k}^{x}\delta_{w}^{l}\delta_{y}^{n}=\frac{4\Lambda}{D-2}\delta_{w}^{l}\delta_{y}^{n}\delta_{k}^{x}\delta_{m}^{z},
\end{equation}
 
\begin{align}
\left[\frac{\partial F}{\partial R_{j}^{i}\partial R_{ln}^{km}}\right]_{\bar{R}} & =\delta_{i}^{n}\bar{R}_{km}^{lj}+\delta_{m}^{j}\bar{R}_{ki}^{ln}\nonumber \\
 & =\frac{2\Lambda}{\left(D-1\right)\left(D-2\right)}\delta_{i}^{n}\left(\delta_{k}^{l}\delta_{m}^{j}-\delta_{m}^{l}\delta_{k}^{j}\right)+\frac{2\Lambda}{\left(D-1\right)\left(D-2\right)}\delta_{m}^{j}\left(\delta_{k}^{l}\delta_{i}^{n}-\delta_{i}^{l}\delta_{k}^{n}\right),
\end{align}
 summing up to 
\begin{align}
f_{\text{quad-equal}}\left(R_{b}^{a},R_{cd}^{ab}\right)= & \frac{16D\Lambda^{3}}{\left(D-1\right)\left(D-2\right)^{3}}-\frac{24\Lambda^{2}}{\left(D-1\right)\left(D-2\right)^{2}}R\nonumber \\
 & +\frac{2\Lambda}{D-2}R_{abcd}^{2}+\frac{8\Lambda}{\left(D-1\right)\left(D-2\right)}R_{ab}^{2}.
\end{align}

\subsubsection{The term $RR_{rs}^{pq}R_{pq}^{rs}$:}

Let $F\left(R,R_{cd}^{ab}\right)\equiv RR_{rs}^{pq}R_{pq}^{rs}$.
Then, computing the relevant derivatives 
\begin{equation}
F\left(\bar{R},\bar{R}_{cd}^{ab}\right)=\bar{R}\bar{R}^{pqrs}\bar{R}_{pqrs}=\frac{16D^{2}\Lambda^{3}}{\left(D-1\right)\left(D-2\right)^{3}},
\end{equation}
 
\begin{equation}
\left[\frac{\partial F}{\partial R}\right]_{\bar{R}}=\bar{R}_{rs}^{pq}\bar{R}_{pq}^{rs}=\frac{8D\Lambda^{2}}{\left(D-1\right)\left(D-2\right)^{2}},
\end{equation}
 
\begin{equation}
\left[\frac{\partial F}{\partial R_{ln}^{km}}\right]_{\bar{R}}=2\bar{R}\bar{R}_{km}^{ln}=\frac{8D\Lambda^{2}}{\left(D-1\right)\left(D-2\right)^{2}}\left(\delta_{k}^{l}\delta_{m}^{n}-\delta_{m}^{l}\delta_{k}^{n}\right)
\end{equation}
 
\begin{equation}
\left[\frac{\partial^{2}F}{\partial R_{xz}^{wy}\partial R_{ln}^{km}}\right]_{\bar{R}}=2\bar{R}\delta_{w}^{l}\delta_{y}^{n}\delta_{k}^{x}\delta_{m}^{z}=\frac{4D\Lambda}{D-2}\delta_{w}^{l}\delta_{y}^{n}\delta_{k}^{x}\delta_{m}^{z},
\end{equation}
 
\begin{equation}
\left[\frac{\partial F}{\partial R\partial R_{ln}^{km}}\right]_{\bar{R}}=2\bar{R}_{km}^{ln}=\frac{4\Lambda}{\left(D-1\right)\left(D-2\right)}\left(\delta_{k}^{l}\delta_{m}^{n}-\delta_{m}^{l}\delta_{k}^{n}\right),
\end{equation}
 the equivalent quadratic Lagrangian becomes 
\begin{align}
f_{\text{quad-equal}}\left(R,R_{cd}^{ab}\right)= & \frac{16D^{2}\Lambda^{3}}{\left(D-1\right)\left(D-2\right)^{3}}-\frac{24D\Lambda^{2}}{\left(D-1\right)\left(D-2\right)^{2}}R\nonumber \\
 & +\frac{2D\Lambda}{D-2}R_{abcd}^{2}+\frac{8\Lambda}{\left(D-1\right)\left(D-2\right)}R^{2}.
\end{align}

\subsubsection{The term $R_{q}^{p}R_{s}^{r}R_{pr}^{qs}$:}

Let $F\left(R_{b}^{a},R_{cd}^{ab}\right)\equiv R_{q}^{p}R_{s}^{r}R_{pr}^{qs}$.
Then, computing the relevant derivatives 
\begin{equation}
F\left(\bar{R}_{b}^{a},\bar{R}_{cd}^{ab}\right)=\bar{R}_{q}^{p}\bar{R}_{s}^{r}\bar{R}_{pr}^{qs}=\frac{8D\Lambda^{3}}{\left(D-2\right)^{3}},
\end{equation}
 
\begin{equation}
\left[\frac{\partial F}{\partial R_{j}^{i}}\right]_{\bar{R}}=2\bar{R}_{s}^{r}\bar{R}_{ir}^{js}=\frac{8\Lambda^{2}}{\left(D-2\right)^{2}}\delta_{i}^{j},
\end{equation}
 
\begin{equation}
\left[\frac{\partial F}{\partial R_{ln}^{km}}\right]_{\bar{R}}=\bar{R}_{k}^{l}\bar{R}_{m}^{n}=\frac{4\Lambda^{2}}{\left(D-2\right)^{2}}\delta_{k}^{l}\delta_{m}^{n},
\end{equation}
 
\begin{equation}
\left[\frac{\partial^{2}F}{\partial R_{j}^{i}\partial R_{l}^{k}}\right]_{\bar{R}}=2\bar{R}_{ik}^{jl}=\frac{4\Lambda}{\left(D-1\right)\left(D-2\right)}\left(\delta_{i}^{j}\delta_{k}^{l}-\delta_{k}^{j}\delta_{i}^{l}\right),
\end{equation}
 
\begin{equation}
\left[\frac{\partial F}{\partial R_{j}^{i}\partial R_{ln}^{km}}\right]_{\bar{R}}=2\bar{R}_{m}^{n}\delta_{k}^{j}\delta_{i}^{l}=\frac{4\Lambda}{D-2}\delta_{m}^{n}\delta_{k}^{j}\delta_{i}^{l},
\end{equation}
 the equivalent quadratic Lagrangian becomes 
\begin{align}
f_{\text{quad-equal}}\left(R_{b}^{a},R_{cd}^{ab}\right)= & \frac{8D\Lambda^{3}}{\left(D-2\right)^{3}}-\frac{12\Lambda^{2}}{\left(D-2\right)^{2}}R+\frac{2\Lambda}{\left(D-1\right)\left(D-2\right)}R^{2}+\frac{2\left(2D-3\right)\Lambda}{\left(D-1\right)\left(D-2\right)}R_{ab}^{2}.
\end{align}

\end{document}